%% file: sm4.tex
\documentclass[11pt]{article}

\usepackage{graphics}
\usepackage{cite}
\usepackage{rotating}
\usepackage{url}
\usepackage{xspace}

\makeatletter
\if@twoside \oddsidemargin -17pt \evensidemargin 00pt
\else \oddsidemargin 00pt \evensidemargin 00pt
\fi
\expandafter\ifx\csname mathrm\endcsname\relax\def\mathrm#1{{\rm #1}}\fi

\renewcommand{\theequation}{\thesection.\arabic{equation}}
\newcounter{saveeqn}

\@addtoreset{equation}{section}

\makeatother

\def\beq{\begin{equation}}
\def\eeq{\end{equation}}
\def\beqar{\begin{eqnarray}}
\def\eeqar{\end{eqnarray}}
\def\barr#1{\begin{array}{#1}}
\def\earr{\end{array}}
\def\bfi{\begin{figure}}
\def\efi{\end{figure}}
\def\btab{\begin{table}}
\def\etab{\end{table}}
\def\bce{\begin{center}}
\def\ece{\end{center}}

\def\text{\textstyle}
\def\arraystretch{1.0}

\def\refeq#1{\mbox{(\ref{#1})}}
\def\reffi#1{\mbox{Fig.~\ref{#1}}}

\def\refta#1{\mbox{Table~\ref{#1}}}

\def\refse#1{\mbox{Section~\ref{#1}}}

\def\citere#1{\mbox{Ref.~\cite{#1}}}

\newcommand{\GeV}{\unskip\,\mathrm{GeV}}

\newcommand{\TeV}{\unskip\,\mathrm{TeV}}

\def\mathswitch#1{\relax\ifmmode#1\else$#1$\fi}
\def\mathswitchr#1{\relax\ifmmode{\mathrm{#1}}\else$\mathrm{#1}$\fi}
\def\mathswitchit#1{\relax\ifmmode{#1}\else$#1$\fi}

\newcommand{\PW}{\mathswitchr W}
\newcommand{\PZ}{\mathswitchr Z}
\newcommand{\Pg}{\mathswitchr g}
\newcommand{\PH}{\mathswitchr H}
\newcommand{\Pe}{\mathswitchr e}
\newcommand{\Pf}{f}
\newcommand{\Pl}{\mathswitch l}
\newcommand{\Pq}{\mathswitchit q}
\newcommand{\Pep}{\mathswitchr {e^+}}
\newcommand{\Pem}{\mathswitchr {e^-}}
\newcommand{\MW}{\mathswitch {M_{\PW}}}

\newcommand{\MZ}{\mathswitch {M_{\PZ}}}
\newcommand{\MH}{\mathswitch {M_{\PH}}}

\newcommand{\Mt}{\mathswitch {m_{\PQt}}}

\newcommand{\GF}{\mathswitch {G_\mu}}

\newcommand{\alphas}{\alpha_{\mathrm{s}}}

\def\ie{i.e.\ }

\newcommand{\br}{{\mathrm{BR}}}

\newcommand{\EW}{{\mathrm{EW}}}
\newcommand{\QCD}{{\mathrm{QCD}}}

\newcommand{\THU}{{\mathrm{THU}}}

\newcommand{\LO}{{\mathrm{LO}}}
\newcommand{\NLO}{{\mathrm{NLO}}}

\def\Re{\mathop{\mathrm{Re}}\nolimits}

\def\draftdate{\relax}
\def\mda{\relax}
\def\mua{\relax}
\def\mla{\relax}
\def\draft{
\def\thtystars{******************************}
\def\sixtystars{\thtystars\thtystars}
\typeout{}
\typeout{\sixtystars**}
\typeout{* Draft mode!
         For final version remove \protect\draft\space in source file *}
\typeout{\sixtystars**}
\typeout{}
\def\draftdate{\today}
\def\mua{\marginpar[\boldmath\hfil$\uparrow$]%
                   {\boldmath$\uparrow$\hfil}%
                    \typeout{marginpar: $\uparrow$}\ignorespaces}
\def\mda{\marginpar[\boldmath\hfil$\downarrow$]%
                   {\boldmath$\downarrow$\hfil}%
                    \typeout{marginpar: $\downarrow$}\ignorespaces}
\def\mla{\marginpar[\boldmath\hfil$\rightarrow$]%
                   {\boldmath$\leftarrow $\hfil}%
                    \typeout{marginpar: $\leftrightarrow$}\ignorespaces}
\def\Mua{\marginpar[\boldmath\hfil$\Uparrow$]%
                   {\boldmath$\Uparrow$\hfil}%
                    \typeout{marginpar: $\Uparrow$}\ignorespaces}
\def\Mda{\marginpar[\boldmath\hfil$\Downarrow$]%
                   {\boldmath$\Downarrow$\hfil}%
                    \typeout{marginpar: $\Downarrow$}\ignorespaces}
\def\Mla{\marginpar[\boldmath\hfil$\Rightarrow$]%
                   {\boldmath$\Leftarrow $\hfil}%
                    \typeout{marginpar: $\Leftrightarrow$}\ignorespaces}
\overfullrule 5pt
\oddsidemargin -15mm
\marginparwidth 29mm
}

\makeatletter

\def\eqnarray{\stepcounter{equation}\let\@currentlabel=\theequation
\global\@eqnswtrue
\global\@eqcnt\z@\tabskip\@centering\let\\=\@eqncr
$$\halign to \displaywidth\bgroup\hskip\@centering
  $\displaystyle\tabskip\z@{##}$\@eqnsel&\global\@eqcnt\@ne
  \hskip 2\arraycolsep \hfil${##}$\hfil
  &\global\@eqcnt\tw@ \hskip 2\arraycolsep $\displaystyle\tabskip\z@{##}$\hfil
   \tabskip\@centering&\llap{##}\tabskip\z@\cr}
\def\appendix{\par
 \setcounter{section}{0} \setcounter{subsection}{0}
 \def\thesection{\Alph{section}}}

\newcommand{\lsim}
{\;\raisebox{-.3em}{$\stackrel{\displaystyle <}{\sim}$}\;}
\newcommand{\gsim}
{\;\raisebox{-.3em}{$\stackrel{\displaystyle >}{\sim}$}\;}

\def\dsl{\mathpalette\make@slash}
\def\make@slash#1#2{\setbox\z@\hbox{$#1#2$}%
  \hbox to 0pt{\hss$#1/$\hss\kern-\wd0}\box0}

\makeatother

\newcommand{\bmid}{\Bigr|}

\providecommand{\PQu}{\mathswitchr {u}}
\providecommand{\PQd}{\mathswitchr {d}}
\providecommand{\PQc}{\mathswitchr {c}}
\providecommand{\PQs}{\mathswitchr {s}}
\providecommand{\PQb}{\mathswitchr {b}}
\providecommand{\PQt}{\mathswitchr {t}}

\providecommand{\PQbpr}{\mathswitchr {b'}}
\providecommand{\PQtpr}{\mathswitchr {t'}}
\providecommand{\Plpr}{\mathswitchr {l'}}
\providecommand{\PGnlpr}{\nu_{\Plpr}}
\providecommand{\PGg}{\gamma}
\providecommand{\PGm}{\mu}
\providecommand{\PGmp}{\mu^+}
\providecommand{\PGmm}{\mu^-}
\providecommand{\PGt}{\tau}
\providecommand{\PGne}{\nu_{\Pe}}
\providecommand{\PGnGm}{\nu_{\PGm}}
\providecommand{\PGnGt}{\nu_{\PGt}}

\newcommand{\Mbp}{\mathswitch {m_{\PQbpr}}}
\newcommand{\Mtp}{\mathswitch {m_{\PQtpr}}}
\newcommand{\Mlp}{\mathswitch {m_{\Plpr}}}
\newcommand{\Mnp}{\mathswitch {m_{\PGnlpr}}}
\newcommand{\Mfp}{\mathswitch {m_{\Pf'}}}

\newcommand{\MGfpr}{\mathswitch {m_{f'}}}

\newcommand{\MGqpr}{\mathswitch {m_{\ssQ}}}
\newcommand{\MGlpr}{\mathswitch {m_{\ssL}}}

\newcommand{\Bref}[1]{Ref.~\cite{#1}}
\newcommand{\Brefs}[1]{Refs.~\cite{#1}}

\newcommand{\NNLO}{{\mathrm{NNLO}}}
\newcommand{\mySM}{\rm{\scriptscriptstyle{SM}}}

\newcommand{\myF}{\rm{\scriptscriptstyle{F}}}
\newcommand{\myEW}{{\rm\scriptscriptstyle{E}\scriptscriptstyle{W}}}

\newcommand{\ssQ}{{\scriptscriptstyle{Q}}}

\newcommand{\ssL}{{\scriptscriptstyle{L}}}

\newcommand{\eqn}[1]{Eq.~(\ref{#1})}

\newcommand{\lpar}{\left(}                            
\newcommand{\rpar}{\right)}

\providecommand{\Hbb}{\PH \to \PQb \bar{\PQb}}
\providecommand{\Htautau}{\PH \to \tau^+\tau^-}
\providecommand{\Hmumu}{\PH \to \mu^+\mu^-}
\providecommand{\Hss}{\PH \to \PQs \bar{\PQs}}
\providecommand{\Hcc}{\PH \to \PQc \bar{\PQc}}
\providecommand{\Htt}{\PH \to \PQt \bar{\PQt}}
\providecommand{\Hgg}{\PH \to \Pg\Pg}

\providecommand{\HZga}{\PH \to \PZ\gamma}
\providecommand{\HWW}{\PH \to \PW\PW}
\providecommand{\HZZ}{\PH \to \PZ\PZ}

\begin{document}

\thispagestyle{empty}
\def\thefootnote{\fnsymbol{footnote}}
\setcounter{footnote}{1}
\null
\draftdate\hfill  FR-PHENO-2011-022 \\
\strut\hfill MPP-2011-135 \\
\strut\hfill PSI-PR-11-03 \\
\strut\hfill TTK-11-57 \\[1em]
\centerline{\bf LHC Higgs Cross Section Working Group}
\vskip 1cm
\begin{center}
  {\Large \boldmath{\bf Higgs production and decay \\[.4em]
  with a fourth Standard-Model-like fermion generation}
\par} \vskip 2.5em
{\large
{\sc A.~Denner$^{1}$,
S.~Dittmaier$^{2}$,
A.~M\"uck$^{3}$,
G.~Passarino$^{4}$, \\[.3em]
M.~Spira$^{5}$,
C.~Sturm$^{6}$,
S.~Uccirati$^{1}$
and M.M.~Weber$^{6}$
}\\[3ex]
{\normalsize \it
$^1$Universit\"at W\"urzburg, Institut f\"ur Theoretische Physik und Astrophysik, \\
D-97074 W\"urzburg, Germany
}\\[1ex]
{\normalsize \it
$^2$Albert-Ludwigs-Universit\"at Freiburg, Physikalisches Institut, \\
D-79104 Freiburg, Germany
}\\[1ex]
{\normalsize \it
$^3$Institut f\"ur Theoretische Teilchenphysik und Kosmologie, RWTH
Aachen,
\\
D-52056 Aachen, Germany}
\\[1ex]
{\normalsize \it
$^4$Dipartimento di Fisica Teorica, Universit\`a
di Torino, Italy \\
INFN, Sezione di Torino, Italy}
\\[1ex]
{\normalsize \it
$^5$Paul Scherrer Institut, W\"urenlingen und Villigen,\\
CH-5232 Villigen PSI, Switzerland}
\\[1ex]
{\normalsize \it
$^6$Max-Planck-Institut f\"ur Physik (Werner-Heisenberg-Institut)\\
D-80805 M\"unchen, Germany}
}
\par \vskip 1em
\end{center}\par
\vfill \vskip .0cm \vfill {\bf Abstract:} \par State-of-the-art
predictions for the Higgs-boson production cross section via gluon
fusion and for all relevant Higgs-boson decay channels are presented
in the presence of a fourth Standard-Model-like fermion generation.
The qualitative features of the most important differences to the
genuine Standard Model are pointed out, and the use of the available
tools for the predictions is described. For a generic mass scale of
$400{-}600\GeV$ in the fourth generation explicit numerical results for the
cross section and decay widths are presented, revealing extremely
large electroweak radiative corrections, e.g., to the cross section
and the Higgs decay into WW or ZZ pairs, where they amount to about
$-50\%$ or more.  This signals the onset of a non-perturbative regime
due to the large Yukawa couplings in the fourth generation. An
estimate of the respective large theoretical uncertainties is
presented as well.
\par
\vskip 1cm
\noindent
March 2012
\par
\null
\setcounter{page}{0}
\clearpage
\def\thefootnote{\arabic{footnote}}
\setcounter{footnote}{0}

\section{Introduction}

In the last years intensive studies at the LHC aimed at putting
exclusion limits on an extension of the Standard Model (SM) with an
additional fourth generation of heavy fermions. Besides direct
searches for heavy quarks \cite{Chatrchyan:2011em,Luk:2011np}, Higgs
production in gluon fusion ($\Pg\Pg$-fusion) is an important channel
in this respect \cite{Aad:2011qi,CMShiggs:2011}, as it is particularly
sensitive to new coloured, heavy particles.%
\footnote{Results of similar searches at Tevatron can be found in
  \Brefs{Aaltonen:2011tq,Aaltonen:2011vr} and \Bref{Aaltonen:2010sv},
  respectively.}  Given the spectacular modification in the
Higgs-boson cross section at hadron colliders that can be tested
easily with LHC data, a SM with a fourth generation of heavy fermions
stimulates great interest.

So far, the experimental analysis has concentrated on models with
ultra-heavy fourth-genera\-ti\-on fermions, excluding the possibility
that the Higgs boson decays to heavy neutrinos. Furthermore, in the
literature \cite{Anastasiou:2011qw} the two-loop electroweak
corrections to $\Pg\Pg$-fusion have been included only under the
assumption that they are dominated by light fermions. At the moment,
however, the experimental strategy consists in computing the ratio of
Higgs-production cross sections in the SM with a 4th generation of
fermions (SM4) and the SM with $3$ generations (SM3), $R = \sigma({\rm
  SM4})/\sigma({\rm SM3})$, with {\sc HIGLU}~\cite{Spira:1995mt} while
all next-to-leading-order (NLO) electroweak (EW) radiative corrections
are switched off.  The experimental situation is as follows: the
search in all channels, updated for the \textit{International
  Europhysics Conference on High Energy Physics 2011 (HEP2011)} and
the \textit{XXV International Symposium on Lepton Photon Interactions
  at High Energies (LP11)}, requires Higgs-boson masses $\MH <
120\GeV$ or $\MH > 600\GeV$ (ATLAS and CMS ex-aequo
\cite{cerntalk2011}).  At low $\MH$, LHC limits are more stringent
than Tevatron limits.  However, in all existing analyses complete NLO
EW corrections are not included.  Therefore, changes of up to $10\GeV$
are expected in limits at the low end while changes of the order of
$30\GeV$ are possible in the high-mass region \cite{cerntalk2011}.

Leading-order (LO) or NLO QCD predictions typically depend only weakly
on the precise values of the masses of the heavy fermions and approach
a constant value in the limit of very heavy fermion masses.  In
contrast, NLO EW corrections are enhanced by powers of the masses of
the heavy fermions and thus induce a strong dependence of the results
on these masses and a breakdown of perturbation theory for very heavy
fermions.

While the complete electroweak corrections to Higgs production in SM4
at the LHC have already been calculated in \Bref{Passarino:2011kv}, we
present in this paper for the first time results for all relevant
Higgs-boson decay channels including NLO electroweak corrections in
SM4. For ultra-heavy fermions the leading corrections can be obtained
easily within an effective theory \cite{Chanowitz:1978uj}. However,
for heavy fermions with masses at the level of $500\GeV$ the
asymptotic results are not precise enough and in particular for a
heavy Higgs boson they are not valid. Including the complete NLO
corrections, we discuss the corresponding predictions for various
scenarios of heavy fermion masses and provide estimates of the
theoretical uncertainties.

The paper is organized as follows: In \refse{sec:setup} we define our
general setup. In \refse{sec:Hprod} we describe the calculation of the
SM4 contributions to Higgs-boson production via gluon fusion and in
\refse{sec:Hdec} those for Higgs-boson decays into 4 fermions, fermion
pairs, gluon pairs, photon pairs and photon plus $\PZ$~boson. In
\refse{sec:numres} we present numerical results, and \refse{sec:concl}
contains our conclusions.

\section{General setup}
\label{sec:setup}

We study the extension of the SM that includes a 4th generation of
heavy fermions, consisting of an up- and a down-type quark
$(\PQtpr,\PQbpr)$, a charged lepton $(\Plpr)$, and a massive neutrino
$(\PGnlpr)$.  The 4th-generation fermions all have identical gauge
couplings as their SM copies and equivalent Yukawa couplings
proportional to their masses, but are assumed not to mix with the
other three SM generations.

\begin{sloppypar}
Experimentally, 4th-generation fermions are strongly constrained.
Direct experimental searches from the Tevatron
\cite{Aaltonen:2011tq,Aaltonen:2011vr} and the LHC
\cite{Chatrchyan:2011em,Luk:2011np} yield lower limits, in particular
on the masses of the heavy quarks:
\beq
\Mbp > 361\GeV, \qquad \Mtp> 450\GeV\quad \mathrm{at}\ 95\% \mathrm{CL}. 
\eeq
Stringent bounds on the mass splittings of the heavy fermions result
from electroweak precision data \cite{Kribs:2007nz}, more precisely
from experimental constraints on the $S$ and $T$ parameters of Peskin
and Takeuchi \cite{Peskin:1991sw}. These constraints typically require
mass splittings for the heavy quarks and leptons. Nevertheless also a
mass-degenerate 4th family is not excluded if one allows for flavour
mixing of the 4th-generation fermions \cite{Eberhardt:2010bm}.  While
4th-generation models can accomodate a heavier Higgs boson as the SM3,
very large values of a SM-like Higgs boson are not favoured
\cite{Erler:2010sk}.
Since the Yukawa couplings of the heavy fermions are proportional to
their masses, perturbation theory breaks down for masses of the heavy
fermions above $\sim500\GeV$ \cite{Chanowitz:1978uj}.  In the presence
of heavy fermions, non-perturbative analyses on the lattice push the
allowed Higgs masses to larger values \cite{Gerhold:2010wv}.
\end{sloppypar}

The main goal of this paper is to provide the electroweak corrections
within SM4 for Higgs production and decay.  Owing to screening (see
\refse{sec:Hprod}), LO or NLO QCD predictions
typically depend only weakly on the precise values of the masses of
the heavy fermions.  Therefore, experimental analyses used very heavy
masses for the extra fermions in order to derive conservative limits.
When complete NLO EW corrections are included, the situation changes
dramatically.  Since the NLO EW corrections are enhanced by powers of
the masses of the heavy fermions, perturbation theory breaks down for
fermion masses above $\sim500\GeV$ and perturbative results become
questionable.  Therefore, we focus on 4th-generation masses between
400 and $550\GeV$, i.e.\ values above the direct search bounds but
small enough for perturbation theory to be still viable, and study
different scenarios that are in agreement with electroweak precision
tests. In detail, we consider scenarios that are consistent with the
constraints derived in \citere{Baak:2011ze} (see in particular Figure
13). We choose
\beq\label{eq:scenarios_a}
  \Mtp = 500\GeV, \qquad \Mlp=450\GeV 
\eeq
and consider three different mass splittings for heavy quarks and
leptons each for three values of the Higgs-boson mass:
\beq\label{eq:scenarios_b}
\arraycolsep 7pt
\renewcommand{\arraystretch}{1.3}
\begin{array}{c|c|c|c}
\label{tab:massspectrum}
\MH\, [\mathrm{GeV}]            & 120 & 350 & 600 \\       
\hline
\Mtp-\Mbp\, [\mathrm{GeV}]      & -50,~~0,~{+50} &-50,~~0,~{+50} &-50,~~0,~{+50} \\
\hline
 \Mnp-\Mlp\, [\mathrm{GeV}]    & -100, -75, -50 &-100, -75, -50 &-150, -100, -50 
\end{array}
\eeq
Moreover, we provide a scan over Higgs-boson masses from $100\GeV$ to
$600\GeV$ for the scenario 
\beq\label{eq:scenarios_c}
\Mtp = 500\GeV, \qquad \Mlp=450\GeV, \qquad \Mbp=450\GeV, \qquad \Mnp=375\GeV,
\eeq
which is a particular case of \refeq{eq:scenarios_a}/\refeq{eq:scenarios_b}.
Note that for this range of Higgs-boson masses, the decay of the Higgs
boson into a pair of heavy fermions is kinematically not allowed in
the scenarios considered above.

In addition, we provide results for the extreme scenario
\beqar
  \Mbp &=& \Mlp= \Mnp = 600\GeV, \nonumber \\
  \Mtp &=& \Mbp + \left[1 + \frac{1}{5} \ln\biggl(\frac{\MH}{115\GeV}\biggr)\right] \, 50 \GeV,
\label{eq:masses}
\eeqar
where the relation among the heavy fermion masses is used to avoid
current exclusion limits from EW precision data (see
\Bref{Kribs:2007nz}).  This setup is at the border between the
perturbative and the non-perturbative regime.  It
is as close as possible to the infinite 4th-generation case, which was
used by ATLAS and CMS to get conservative exclusion limits, and in fact
was employed to derive experimental limits on the Higgs-boson mass
within SM4\cite{CMShiggs:2011}.

In the extreme scenario \refeq{eq:masses}, we give results for Higgs
masses between $100\GeV$ and $1\TeV$ for an on-shell Higgs boson. For
Higgs masses above $\sim500\GeV$, the off-shellness of the Higgs boson
becomes relevant, and finite-width effects and background
contributions can become important.  A treatment of these effects is
very difficult and beyond the scope of the present paper. Attempts to
describe these effects in the SM can be found in
\Brefs{Anastasiou:2011pi,Goria:2011wa} and a discussion of the
corresponging theoretical uncertainties in \Bref{Giampiero_webpage}.

\section{Higgs-boson production via gluon fusion \label{sec:Hprod}}

In the Standard Model with three fermion generations the Higgs-boson
production via gluon fusion is basically determined at leading order by
the contribution of just the one-loop diagram where a top quark is
running in the loop (the bottom-quark loop can be neglected in a first
approximation).  Despite the presence of the Yukawa coupling
proportional to the top-quark mass, the LO amplitude goes at high $\Mt$
asymptotically towards a constant (screening).  Moving from SM3 to SM4,
the LO $\Pg\Pg$-fusion cross section for a light Higgs boson is then
about nine times larger than the one of SM3, because three heavy
fermions instead of one propagate in the loop~\cite{Georgi:1977gs}.

The screening behaviour at leading order is preserved by QCD corrections
\cite{Spira:1995rr,Anastasiou:2011qw}.  Concerning the EW corrections
the leading behaviour for high values of the masses in the fourth
generation is known since long~\cite{Djouadi:1994ge,Djouadi:1997rj} (see
also \Bref{Fugel:2004ug}) and shows an enhancement of radiative
corrections proportional to the square of the (heavy) fermion masses.
This enhancement is, however, accidentally spoiled in the quark sector
in presence of degenerate $\PQtpr{-}\PQbpr$ quarks, while it still survives in
the (heavy) lepton sector.  Recently the complete two-loop EW
corrections to Higgs-boson production through $\Pg\Pg$-fusion at the LHC
in SM4 have been computed in \Bref{Passarino:2011kv} by extending the
corresponding calculations of \Brefs{Actis:2008ts,Actis:2008ug} in SM3.
In~\Bref{Passarino:2011kv} explicit results have been given in the 
scenario \refeq{eq:masses} of large fourth-generation masses; in this section we
determine the complete two-loop EW corrections using the same methods,
however, for different mass scenarios.

\begin{figure}
\begin{center}
\includegraphics[height=9cm,bb=0 0 567 384]
{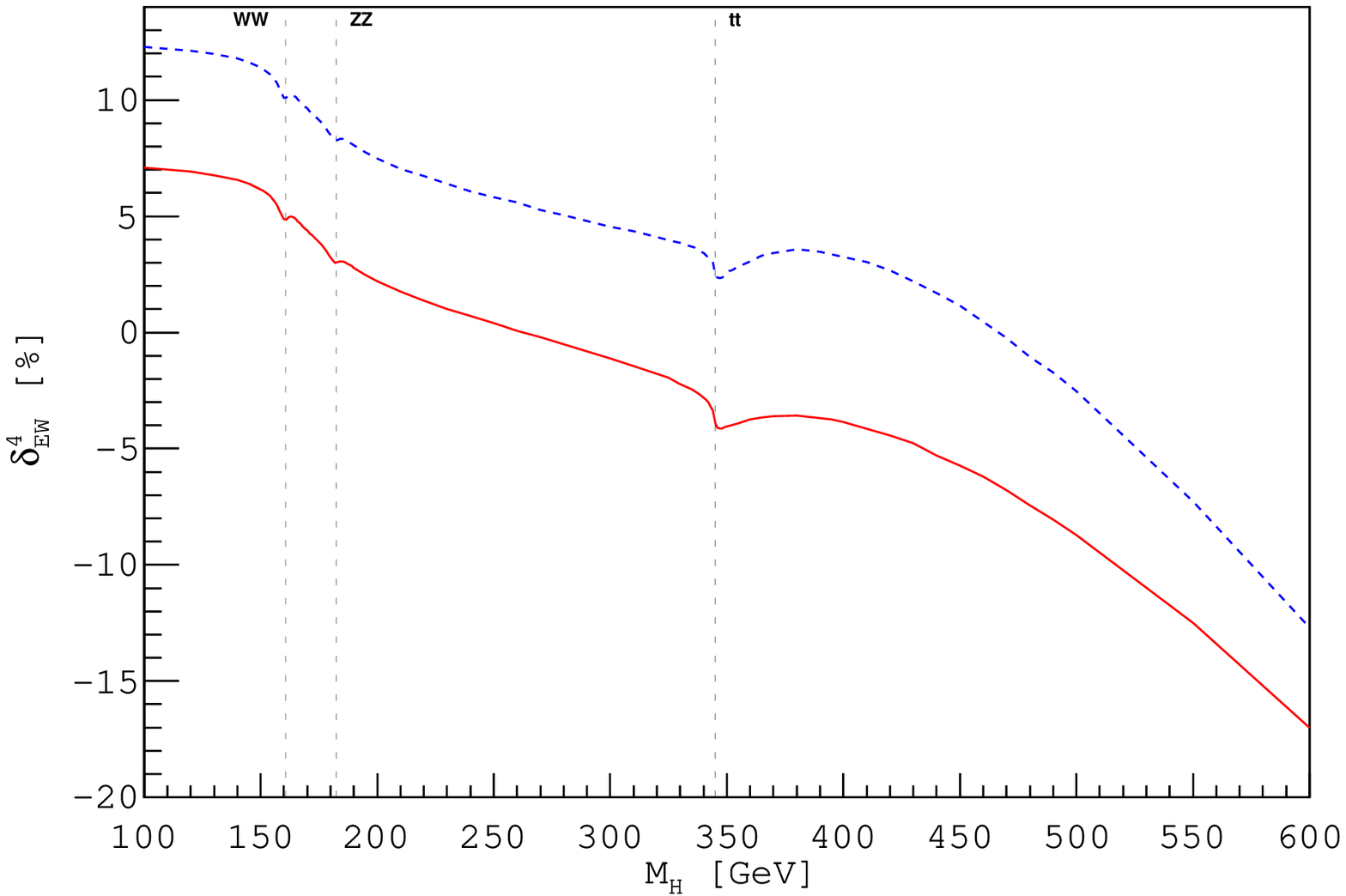}
\end{center}
\vspace{-1em}
\caption[]{\label{SM4_common_1} Relative corrections in SM4 due to
  two-loop EW corrections to $\Pg \Pg \to \PH$.  The solid, red curve
  corresponds to the mass scenario $\Mtp = 500\GeV$, $\Mbp=450\GeV$,
  $\Mnp=375\GeV$, $\Mlp=450\GeV$, while the dashed, blue curve
  corresponds to the extreme scenario of \eqn{eq:masses}.  }
\end{figure}

Let us start with the scan over Higgs-boson masses specified in 
\eqn{eq:scenarios_c} of \refse{sec:setup}.
The relative EW two-loop correction $\delta^{(4)}_{\myEW}$
with respect to the leading-order
cross section $\sigma^{\LO}_{\mySM 4}\lpar \Pg\Pg \to \PH\rpar$ in SM4
are defined via the corrected cross section by
\beq
\sigma_{\mySM 4}\lpar \Pg\Pg \to \PH\rpar =
\sigma^{\LO}_{\mySM 4}\lpar \Pg\Pg \to \PH\rpar\,
\Bigl( 1 + \delta^{(4)}_{\myEW}\Bigr)~.
\eeq
The result for $\delta^{(4)}_{\myEW}$ in this scenario is shown in
\reffi{SM4_common_1} (solid, red curve).  The vertical lines in the
figure denote the location of the $\PW\PW$-, $\PZ\PZ$-, and
$\PQt\bar{\PQt}$-thresholds.  The NLO EW corrections due to the fourth
generation are positive for a light Higgs-boson mass and start to become
negative for Higgs-boson masses above $260\GeV$. 
Figure~\ref{SM4_common_1} also shows the behaviour of
$\delta^{(4)}_{\myEW}$ in the extreme scenario of \eqn{eq:masses}
(dashed, blue curve), which can be considered as the upper bound of EW
corrections in the perturbative regime.  Some values of the solid, red curve
of~\reffi{SM4_common_1} are also listed in~\refta{tab:EWNLO}.
\begin{table}
\begin{center}
\caption[]{\label{tab:EWNLO} Relative NLO EW corrections to the
  $\Pg\Pg\to \PH$ cross sections in SM4, for the mass scenario
  $\Mtp=500\GeV$, $\Mbp=450\GeV$, $\Mnp=375\GeV$, $\Mlp=450\GeV$.  The
  absolute numerical integration error is well below $0.01\%$ for
  Higgs-boson masses below the $\PQt\bar{\PQt}$-threshold and below $0.05\%$
  above it.  }
\vspace{1em}
\input{plots/SM4EW_ctab}

\end{center}
\end{table}

In addition to these scenarios for the masses of the fourth generation
of fermions, we have also performed a scan in the $\Mbp{-}\Mnp$ space as
given in \eqn{tab:massspectrum} 
for fixed values of the masses $\Mtp=500 \GeV$, $\Mlp=450 \GeV$ and for
three values of the Higgs-boson mass $\MH$=120, 350, 600 $\GeV$.  These
results for the relative correction are listed
in~\refta{tab:EWNLOscan}.
\begin{table}
\begin{center}
\caption[]{\label{tab:EWNLOscan} Relative NLO EW corrections to the
  $\Pg\Pg\to \PH$ cross sections in SM4 for three different values of
  the Higgs-boson mass $\MH$ with fixed values for the masses $\Mtp=500
  \GeV$, $\Mlp=450 \GeV$ and different values for the masses $\Mbp$,
  $\Mnp$.  The absolute numerical integration error is well below
  $0.002\%$ for $\MH=120\GeV$ and below $0.05\%$ for the other
  Higgs-boson masses.  }
\vspace{1em}
\input{plots/SM4EW_ctabscan}

\end{center}
\end{table}

For the mass scenario of Eq.~\refeq{eq:masses} the EW NLO corrections
become $-100\%$ just before the heavy-quark thresholds of the 4th
generation and also for the mass scenario \refeq{eq:scenarios_c} the
EW NLO corrections become sizable when approaching the heavy-quark
thresholds making in both cases the use of the perturbative approach
questionable.  In the high-mass region we have no solid argument to
estimate the remaining uncertainty and prefer to state that SM4 is in
a fully non-perturbative regime which should be approached with
extreme caution.  For the low-mass region we can do no more than make
educated guesses, based on the expected asymptotic behaviour for a
heavy fourth generation.  At EW NNLO there are diagrams with five
Yukawa couplings; we can therefore expect an enhancement at 3~loops
which goes as the fourth power of the heavy-fermion mass, unless some
accidental screening occurs.  Therefore, assuming a quartic leading
3-loop behaviour in the heavy fermion mass $\Mfp$, we estimate the
remaining uncertainty to be of the order of
$(\alpha/\pi)^2(\Mfp/\MW)^4$ and thus $\sim2\,\%$ for the scenario
\refeq{eq:masses} in the interval
$\MH = 100{-}600\GeV$, even less for the scenarios
\refeq{eq:scenarios_a}/\refeq{eq:scenarios_b}.

Having computed the EW corrections $\delta^{(4)}_{\myEW}$ we should
discuss some aspects of their inclusion in the production cross section
$\sigma\lpar \Pg\Pg \to \PH + X\rpar$, i.e.\ their interplay with QCD
corrections and the remaining theoretical uncertainty. The most accepted
choice is given by
\beq
\sigma^{\myF} = \sigma^{\LO}
\,\Bigl( 1 + \delta_{\QCD}\Bigr)\,\,\Bigl( 1 + \delta_{\myEW}\Bigr)~,
\eeq
which assumes complete factorization of QCD and EW corrections. The
latter is based on the work of \Bref{Anastasiou:2008tj} where it is
shown that, at zero Higgs momentum, exact factorization is violated
but with a negligible numerical impact; the result of
\Bref{Anastasiou:2008tj} can be understood in terms of soft-gluon
dominance. The residual part beyond the soft-gluon-dominated part
contributes up to $5{-}10\%$ to the total inclusive cross section (for
Higgs-boson masses up to $1\TeV$).  Since the EW corrections are less
than 50\% in the considered Higgs-mass range, the non-factorizable
effects of EW corrections should be below $5\%$ in SM4.

\section{Higgs-boson decays}
\label{sec:Hdec}

\subsection{\boldmath{NLO corrections to $\PH\to 4\Pf$ in SM4}}
\label{se:H4f}
\newcommand{\pfourf}{{\sc Prophecy4f} }

\begin{sloppypar}
The results of the $\PH\to 4\Pf$ decay channels have been
  obtained using the Monte Carlo generator
  \pfourf \cite{Prophecy4f,Bredenstein:2006rh,Bredenstein:2007ec}
  which has been extended to support the SM4.  \pfourf can
  calculate the EW and QCD NLO corrections to the partial widths for
  all 4f final states, i.e.\ leptonic, semi-leptonic, and hadronic
  final states.  Since the vector bosons are kept off shell, the
  results are valid for Higgs masses below, near, and above the
  on-shell gauge-boson production thresholds. Moreover, all
  interferences between WW and ZZ intermediate states are included at
  LO and NLO.
\end{sloppypar}

The additional corrections in SM4 arise from 4th-generation fermion
loops in the HWW/HZZ vertices, the gauge-boson self-energies, and the
renormalization constants.  For the large 4th-generation masses of
${\cal O}(400{-}600 \GeV)$ considered here, the 4th-generation Yukawa
couplings are large, and the total corrections are dominated by the
4th-generation corrections.  Numerically the NLO corrections amount to
about $-50\%$ for the scenarios
\refeq{eq:scenarios_a}/\refeq{eq:scenarios_b} and $-85\%$ for the
extreme scenario \refeq{eq:masses} and depend only weakly on the
Higgs-boson mass for not too large $\MH$.
The corrections from the 4th generation are taken into account at NLO
with their full mass dependence, but their behaviour for large masses
can be approximated well by the dominant corrections in the
heavy-fermion limit. In this limit the leading contribution can be
absorbed into effective HWW/HZZ interactions in the $\GF$
renormalization scheme via the Lagrangian
\beq
{\cal L}_{HVV} = \sqrt{\sqrt{2} \GF} H
\left[ 2 \MW^2 W^\dagger_\mu W^\mu (1 + \delta_{\PW}^{\mathrm{tot}}) + \MZ^2 Z_\mu Z^\mu (1 + \delta_{\PZ}^{\mathrm{tot}}) \right],
\eeq
where $W,Z,H$ denote the fields for the $\PW$, $\PZ$, and Higgs
bosons.  The higher-order corrections are contained in the factors
$\delta_V^{\mathrm{tot}}$ whose expansion up to two-loop order is
given by
\beq
\delta_V^{\mathrm{tot}(1)} = \delta_u^{(1)} + \delta_V^{(1)}, \qquad
\delta_V^{\mathrm{tot}(2)} = \delta_u^{(2)} + \delta_V^{(2)} + \delta_u^{(1)} \delta_V^{(1)}.
\eeq
The one-loop expressions for a single $\mathrm{SU}(2)$ doublet of heavy
fermions with masses $m_A$, $m_B$ read \cite{Chanowitz:1978uj}
\beq
\delta_u^{(1)} = N_c X_A \left[ \frac{7}{6} (1 + x) + \frac{x}{1-x} \ln x\right], \qquad
\delta_V^{(1)} = - 2 N_c X_A (1+x),
\label{eq:delta1}
\eeq
where $x = m_B^2 / m_A^2$,  $X_A = \GF m_A^2/(8 \sqrt{2} \pi^2)$, and
$N_c=3$ or $1$ for quarks or leptons, respectively.
The results for the two-loop corrections $\delta_V^{\mathrm{tot}(2)}$ can
be found in \Bref{Kniehl:1995ra} for the QCD corrections of ${\cal
  O}(\alphas \GF \Mfp^2)$ and in \Bref{Djouadi:1997rj} for the EW
corrections of ${\cal O}(\GF^2\Mfp^4)$.  The corrected partial
decay width $\Gamma$ is then given by
\beq
\Gamma_{\mathrm{NLO}} \;\approx\; \Gamma_{\mathrm{LO}} \left[1 + \delta_\Gamma^{(1)} +
\delta_\Gamma^{(2)}\right] \;=\;
\Gamma_{\mathrm{LO}} \left[1 + 2 \delta_V^{\mathrm{tot}(1)} +
(\delta_V^{\mathrm{tot}(1)})^2 + 2 \delta_V^{\mathrm{tot}(2)}\right].
\eeq

The size of the two-loop corrections $\delta_\Gamma^{(2)}$ is about
$+(6{-}9)\%$ for the scenarios \refeq{eq:scenarios_a}/\refeq{eq:scenarios_b}
and $+15\%$ for the extreme scenario \refeq{eq:masses} depending only
very weakly on the Higgs mass. Due to the large one-loop corrections
\pfourf includes the two-loop QCD and EW corrections in the
heavy-fermion limit in addition to the exact one-loop corrections.
Although the asymptotic two-loop corrections are not directly
applicable for a heavy Higgs boson, they can be viewed as a
qualitative estimate of the two-loop effects. One should keep in mind
that for a Higgs boson heavier than about $600\GeV$ many more
uncertainties arise owing to the breakdown of perturbation theory.

The leading two-loop terms can be taken as an estimate of the error
from unknown higher-order corrections. This implies an error relative
to the LO of $7\%$ for the scenarios
\refeq{eq:scenarios_a}/\refeq{eq:scenarios_b} and $15\%$ for the
extreme scenario \refeq{eq:masses} on the partial width for all
$\PH\to 4\Pf$ decay channels.  Assuming a scaling law of this error
proportional to $X_A^2$, the uncertainty for general mass scenarios
can be estimated to about $100X_A^2$ relative to the LO prediction.
However, since the correction grows large and negative, the relative
uncertainty on the corrected width gets enhanced to
$100X_A^2/(1-64X_A/3+100X_A^2)$, where the linear term in $X_A$
parametrizes the leading one-loop correction. For the mass $m_A$ in
$X_A$ either the weighted squared average
$m_A^2=N_c(\Mbp^2+\Mtp^2)+\Mlp^2+\Mnp^2$ or the maximal mass
$m_A=\max(\Mbp,\Mtp,\Mlp,\Mnp)$ should be used.  For $\Mfp=500\GeV$
and $\Mfp=600\GeV$ this results roughly in an uncertainty of $ 14\%$
and $50\%$, respectively, on the corrected $\PH\to 4\Pf$ decay widths.

\subsection{\boldmath{$\PH\to \Pf\bar{\Pf}$}}

The decay widths for $\PH\to \Pf\bar{\Pf}$ are calculated with {\sc
  HDECAY} \cite{Djouadi:1997yw} which includes the approximate NLO and
NNLO EW corrections for the decay channels into SM3 fermion pairs in
the heavy-SM4-fermion limit according to \Bref{Djouadi:1997rj} and
mixed NNLO EW/QCD corrections according to \Bref{Kniehl:1995ra}. These
corrections originate from the wave-function renormalization of the
Higgs boson and are thus universal for all fermion species. The
leading one-loop part is given by $\delta_u^{(1)}$ of \eqn{eq:delta1}.
Numerically the EW one-loop correction to the partial decay widths
into fermion pairs amounts to about $+25\%$ or $+40\%$, for the
scenarios \refeq{eq:scenarios_a}/\refeq{eq:scenarios_b} or
\refeq{eq:masses}, respectively, while the two-loop EW and QCD
correction contributes an additional $+5\%$ or $+20\%$.  The
corrections are assumed to factorize from whatever is included in {\sc
  HDECAY}, since the approximate expressions emerge as corrections to
the effective Lagrangian after integrating out the heavy fermion
species.  Thus, {\sc HDECAY} multiplies the relative SM4 corrections
with the full corrected SM3 result including QCD and approximate EW
corrections.  The scale of the strong coupling $\alphas$ has been
identified with the average mass of the heavy quarks $\PQtpr, \PQbpr$
of the 4th generation.

The unknown higher-order corrections from heavy fermions can be
estimated as for the decay $\PH\to4f$ above from the size of the
leading two-loop corrections. Since the corrections enhance the LO
prediction, the uncertainty relative to the corrected width, which we
estimate as $100X_A^2/(1+32X_A/3+100X_A^2)$ is reduced, resulting in a
theoretical uncertainty for the SM4 part to the full partial decay
widths into fermion pairs of $5\%$ and $10\%$ for the scenarios
\refeq{eq:scenarios_a}/\refeq{eq:scenarios_b} and \refeq{eq:masses},
respectively.  The uncertainties of the SM3 EW and QCD parts are
negligible with respect to that.

\subsection{\boldmath{$\PH\to \Pg\Pg, \PGg\PGg, \PGg\PZ$}}
For the decay modes $\PH\to \Pg\Pg, \PGg\PGg, \PGg\PZ$, {\sc HDECAY}
\cite{Djouadi:1997yw} is used as well.

For $\PH\to \Pg\Pg$, {\sc HDECAY} includes the NNNLO QCD corrections of
the SM in the limit of a heavy top quark
\cite{Spira:1995rr,Inami:1982xt,Djouadi:1991tka,Chetyrkin:1997iv},
applied to the results including the heavy-quark loops. 
While at NNLO the exact QCD corrections in SM4
\cite{Anastasiou:2011qw} are included in this limit, at NNNLO the relative SM3
corrections are added to the relative NNLO corrections and multiplied
by the LO result including the additional quark loops. Since the
failure of such an approximation is less than 1\% at NNLO, we assume
that at NNNLO it is negligible, i.e.\ much smaller than the
residual QCD scale uncertainty of about 3\%.
In addition the full NLO EW corrections of \refse{sec:Hprod} have been
included in factorized form, since the dominant part of the QCD
corrections emerges from the gluonic contributions on top of the
corrections to the effective Lagrangian in the limit of heavy quarks.
Taking besides the scale uncertainty also the missing quark-mass
dependence at NLO and beyond into account, the total theoretical
uncertainties can be estimated to about 5\%.

{\sc HDECAY} \cite{Djouadi:1997yw} includes the
full NLO QCD corrections to the decay mode $\PH\to \PGg\PGg$ supplemented
by the additional contributions of the 4th-generation quarks and charged
leptons according to \Brefs{Spira:1995rr,Djouadi:1990aj}.

Extending the same techniques used for $\PH \to \Pg\Pg$ in
\Bref{Passarino:2011kv}, we have computed the exact amplitude for $\PH
\to \PGg\PGg$ up to NLO (two-loop level). For phenomenological reasons
we restrict the analysis to the range $\MH\lsim150\GeV$.  The
introduction of EW NLO corrections to this decay requires particular
attention.  If we write the amplitude as
\beq
A = A_{\LO} + X_{\PW}\,A_{\NLO} + X_{\PW}^2\,A_{\NNLO} + \dots,
\qquad X_{\PW}= \frac{\GF\MW^2}{8 \sqrt{2} \pi^2},
\eeq
the usual way to include the NLO EW corrections is
\beq
|A|^2 \sim
|A_{\LO}|^2 + 2\,X_{\PW}\,{\rm Re}\Big[A_{\NLO}\,A_{\LO}^{\dagger}\Big] =
|A_{\LO}|^2\,\Big( 1 + \delta_{_\EW}^{(4)} \Big),
\eeq
with
\beq
\delta_{_\EW}^{(4)} = \frac{2\,X_{\PW}\,{\rm Re}[A_{\NLO}A_{\LO}^{^\dagger}]}{|A_{\LO}|^2}.
\eeq
From the explicit calculation it turns out that in all scenarios taken into 
consideration, $\delta_{_\EW}^{(4)}$ is negative and its absolute value is bigger 
than 1.
Part of the problem is related to the fact that at LO the cancellation between 
the $\PW$ and the fermion loops is stronger in SM4 than in SM3 so that the LO 
result is suppressed more, by about a factor of $2$ at the level of the
amplitude and thus by about a factor of $4$ at the level of the decay width.
Furthermore, the NLO corrections are strongly enhanced for ultra-heavy
fermions in the 4th generation; assuming for instance the mass scenario
of \eqn{eq:masses} for the heavy fermions and a Higgs-boson mass of $100\GeV$ we
get $\delta_{_\EW}^{(4)}=- 319\%$; clearly it does not make sense and one should 
always remember that a badly behaving series should not be used to derive
limits on the parameters, \ie on the heavy-fermion masses.
The scenario \refeq{eq:scenarios_c}
is even more subtle. 

In such a situation, where the LO is suppressed,
a proper estimate of $|A|^2$ must also include the next
term in the expansion, i.e. $X_{\PW}^2|A_\NLO|^2$:
\beq
|A|^2 \sim
|A_{\LO}+X_{\PW}A_{\NLO}|^2 =
|A_{\LO}|^2\,\Big( 1 + \bar{\delta}_{_\EW}^{(4)} \Big),
\quad{\rm with}\quad 
\bar{\delta}_{_\EW}^{(4)} = \frac{|A_{\LO}+X_{\PW}A_{\NLO}|^2}{|A_{\LO}|^2} - 1.
\eeq
We define at the amplitude level the $K\,$-factor
\beq
A_{\LO}+X_{\PW}A_{\NLO} = A_{\LO}\,\Big( 1 - K_{\NLO}\Big).
\eeq
$K_{\NLO}$ is a complex quantity, but the imaginary part of $A_{\LO}$ is small 
and therefore the major part of the NLO correction comes from the real part 
of $K_{\NLO}$, which is positive in both scenarios.
The relation between $\bar{\delta}_{_\EW}^{(4)}$ and $K_{\NLO}$ is:
\beq
\bar{\delta}_{_\EW}^{(4)} = 
{\rm Re}\,[K_{\NLO}]\,\Big({\rm Re}[K_{\NLO}] - 2\Big) + {\rm Im}[K_{\NLO}]^2.
\eeq
For both scenarios ${\rm Re}[K_{\NLO}]$ is decreasing with increasing
Higgs-boson mass.  In the extreme scenario of \eqn{eq:masses}, we have
$1 < {\rm Re}[K_{\NLO}] < 2$, then $\bar{\delta}_{_\EW}^{(4)}$
increases in absolute value with increasing Higgs-boson mass (the
small contribution of ${\rm Im}[K_{\NLO}]$ does not change the
behaviour); in the setup \refeq{eq:scenarios_c} ${\rm Re}[K_{\NLO}]
< 1$ and $\bar{\delta}_{_\EW}^{(4)}$ decreases in absolute value.

Furthermore, ${\rm Re}[K_{\NLO}]$ is close to one and, not only 
$A_{\LO}$ is small but also $A= A_{\LO} + X_{\PW}\,A_{\NLO}$ is small 
(even smaller).
Therefore it turns out, that $\bar{\delta}_{_\EW}^{(4)}$ is large (close to one in 
absolute value) and a description of NLO corrections just based on 
$\bar{\delta}_{_\EW}^{(4)}$ could lead to the conclusion that perturbation 
theory breaks down.
However, this conclusion would be too strong. 
The point is:
\begin{itemize}
\item[a)] 
$A_{\LO}$ is accidentally small, 
\item[b)] 
$X_{\PW}\,A_{\NLO}$ is large as expected, but it is accidentally of the same 
order as $A_{\LO}$ and with opposite sign. 
\end{itemize}
We are facing here the problem of dealing with accidentally small 
quantities and it is hard to give expectations on the convergence of 
perturbation theory.
In our opinion, for this process, the effect of including NLO EW corrections 
is thus better discussed in terms of shifted quantities:
\beq
{\overline A}_{\LO} = A_{\LO} + X_{\PW} A_{\NLO},
\qquad
{\overline A}_{\NLO}= A_{\NNLO}.
\eeq
The idea is to use ${\overline A}_{\LO}$ to define a 2-loop corrected 
decay width 
\beq
{\overline \Gamma}_{\LO} = 
\Gamma_{\LO}\,(1+\bar{\delta}_{_\EW}^{(4)}) = 
\Gamma_{\LO}\,\frac{|A_{\LO}+X_{\PW}A_{\NLO}|^2}{|A_{\LO}|^2},
\label{bGamma}
\eeq 
which represents the best starting point of a perturbative expansion.
In other words, the major part of the NLO corrections emerges from an effective 
Lagrangian in the heavy-particle limit, therefore we should consider them as 
correction to the effective Feynman rules and thus to the amplitude.  

To give an estimation of the theoretical error on the missing higher-order 
corrections, we analyse in more details the situation at NLO and try to 
guess the order of magnitude of ${\overline A}_{\NLO}=A_{\NNLO}$.
Assuming for simplicity $\Mbp= \Mtp = \MGqpr$ and $\Mlp = \Mnp = \MGlpr$, 
the amplitude can be written as
\beq
A = A_{\LO}\,\biggl[1  + X_{\PW}\,\Bigl( C_{\ssQ}\,\frac{\MGqpr^2}{\MW^2} +
C_{\ssL}\,\frac{\MGlpr^2}{\MW^2} + R \Bigr) + {\cal O}(X_{\PW}^2) \biggr],
\eeq
where we have factorized out the leading behaviour in the heavy masses.
The quantities $C_{\ssQ,\ssL}$ and $R$ depend on masses, but go towards a constant 
for high fourth-generation masses. 
In the asymptotic region, $\MH < 2\,\MW \ll \MGqpr, \MGlpr$ we require $R$ to 
be a constant and parametrize the $C\,$-functions as
\beq
C_{\ssQ} = -\,\frac{192}{5}\,\lpar 1 + c_{\ssQ}\,\tau \rpar,
\qquad
C_{\ssL} = -\,\frac{32}{3}\,\lpar 1 + c_{\ssL}\,\tau \rpar,
\eeq
where $C_{\ssQ,\ssL}$ are constant and $\tau = \MH^2/(2\MW)^2$. Note
that for $\tau = R = 0$ this is the leading two-loop behaviour
predicted in \Bref{Djouadi:1997rj} (see also \Bref{Fugel:2004ug} for
the top-dependent contribution which we hide here in $R$). By
performing a fit to our exact result we obtain a good agreement in the
asymptotic region, showing that the additional corrections
proportional to $\tau$ play a relevant role. For instance, with
fermions of the 4th generation heavier than $300\GeV$ we have
fit/exact$\,- 1$ less than $5\%$ in the window $\MH = [80{-}130]\GeV$.

Our educated guess for the error estimate is to use the absolute value
of the NLO leading coefficient as the unknown coefficient in the NNLO one,
assuming a leading behaviour of $\MGqpr^4, \MGlpr^4$, \ie no accidental 
cancellations:
\beq
{\overline A}_{\NLO}= A_{\NNLO} \sim 
A_{\LO}\,\bmid C_{\ssQ} + C_{\ssL}\bmid\,\frac{\MGfpr^4}{\MW^4},
\eeq
where we put $\MGfpr={\rm max}(\Mtp,\Mbp,\Mnp,\Mlp)$ in the last term.
In principle one should work at a fixed order in perturbation theory and
estimate the corresponding theoretical uncertainty from the LO--NNLO interference
(since $|A_{\NLO}|^2$ is already part of $|\overline{A}_{\LO}|^2$).
However, the large cancellations in ${\overline A}_{\LO}$ (less relevant in the
conservative scenario) make this option unrealistic and we prefer a
more conservative estimate of the uncertainty, for which we take
\beqar
\bmid A \bmid^2
&\,\sim\;&
\bmid {\overline A}_{\LO} \bmid^2\,
\pm \,2\,X_{\PW}^2 \bmid\Re\big[ {\overline A}_{\NLO}\,{\overline A}_{\LO}^{\dagger}\big]\bmid
\nonumber\\
&\,\sim\;&
\bmid {\overline A}_{\LO} \bmid^2\,
\pm \,2\,X_{\PW}^2\bmid\Re\big[A_{\LO}\,{\overline A}_{\LO}^{\dagger} \big]\bmid\,
|C_{\ssQ} + C_{\ssL}|\,\frac{\MGfpr^4}{\MW^4}.
\label{EWnewderr}
\eeqar
Given our setups the difference between $\Mtp$, $\Mbp$, $\Mnp$ and $\Mlp$ 
is irrelevant in estimating the uncertainty which is now defined as
\beq
\Gamma(\PH \to \PGg\PGg) =
{\overline\Gamma}_{\LO}\,\Big( 1 \pm \delta_{_\THU} \Big),
\qquad\quad
\delta_{_\THU} =
2\,X_{\PW}^2\,\frac{\bmid\Re\big[ {\overline A}^{\dagger}_{\LO}\,A_{\LO}\big]\bmid}
{|{\overline A}_{\LO}|^2}
\bmid C_{\ssQ} + C_{\ssL} \bmid\,\frac{\MGfpr^4}{\MW^4}.
\label{cTHU}
\eeq

The results for the mass scenario \refeq{eq:scenarios_c}
and for the setup of \eqn{eq:masses} are shown 
in \refta{tab:EWNLOGG} and in \refta{tab:EWNLOGGCS}, respectively.
\begin{table}
\begin{center}
\caption[]{\label{tab:EWNLOGG}{
NLO EW corrections to the $\PH \to \PGg\PGg$ decay width (mass scenario of
\eqn{eq:scenarios_c})
according to \eqn{bGamma} and estimate for the missing higher-order corrections
$(\delta_{_\THU})$ relative to ${\overline\Gamma}_{\LO}$ from \eqn{cTHU}.
}}
\vspace{1em}
\begin{tabular}{|c|cccc|}
\hline
\rule[-1ex]{0ex}{3.5ex}%
$\MH\,$[GeV] &
$\Gamma_{\LO}\, [\mathrm{GeV}]$ &
$\bar{\delta}_{\myEW}^{(4)}\,[\%]$  &
${\overline\Gamma}_{\LO}\, [\mathrm{GeV}]$ &
$\delta_{_\THU}\,[\%]$         \\
\hline
\rule[-0ex]{0ex}{2.5ex}%
$100$ & $0.602\cdot10^{-6}$ & $-99.4$ & $0.004\cdot10^{-6}$ & $68.3$ \\
$110$ & $0.938\cdot10^{-6}$ & $-98.2$ & $0.016\cdot10^{-6}$ & $37.1$ \\
$120$ & $1.466\cdot10^{-6}$ & $-96.3$ & $0.054\cdot10^{-6}$ & $23.8$ \\
$130$ & $2.322\cdot10^{-6}$ & $-93.4$ & $0.154\cdot10^{-6}$ & $16.4$ \\
$140$ & $3.802\cdot10^{-6}$ & $-89.2$ & $0.412\cdot10^{-6}$ & $11.6$ \\
$150$ & $6.714\cdot10^{-6}$ & $-83.1$ & $1.133\cdot10^{-6}$ & $ 8.3$
\rule[-1ex]{0ex}{2.5ex}\\
\hline
\end{tabular}
\end{center}
\end{table}
\begin{table}
\begin{center}
\caption[]{\label{tab:EWNLOGGCS}{
NLO EW corrections to the $\PH  \to \PGg\PGg$ decay width (mass scenario of
\eqn{eq:masses}) according to \eqn{bGamma} and estimate for the missing
higher-order ($\delta_{_\THU}$) corrections relative to ${\overline\Gamma}_{\LO}$ from
\eqn{cTHU}.
}}
\vspace{1em}
\begin{tabular}{|c|cccc|}
\hline
\rule[-1ex]{0ex}{3.5ex}%
$\MH\,$[GeV] &
$\Gamma_{\LO}\, [\mathrm{GeV}]$ &
$\bar{\delta}_{\myEW}^{(4)}\,[\%]$  &
${\overline\Gamma}_{\LO}\, [\mathrm{GeV}]$ &
$\delta_{_\THU}\,[\%]$          \\
\hline
\rule[-0ex]{0ex}{2.5ex}%
$100$ & $0.604\cdot10^{-6}$ & $-64.5$ & $0.215\cdot10^{-6}$ & $25.4$ \\
$110$ & $0.942\cdot10^{-6}$ & $-74.4$ & $0.241\cdot10^{-6}$ & $28.2$ \\
$120$ & $1.472\cdot10^{-6}$ & $-83.3$ & $0.246\cdot10^{-6}$ & $32.5$ \\
$130$ & $2.332\cdot10^{-6}$ & $-90.8$ & $0.214\cdot10^{-6}$ & $40.4$ \\
$140$ & $3.820\cdot10^{-6}$ & $-96.6$ & $0.131\cdot10^{-6}$ & $59.7$ \\
$150$ & $6.745\cdot10^{-6}$ & $-99.7$ & $0.020\cdot10^{-6}$ & $>100$
\rule[-1ex]{0ex}{2.5ex}\\
\hline
\end{tabular}
\end{center}
\end{table}
In \refta{tab:EWNLOGG_NS} we show the results at fixed $\MH = 120\GeV$
for different masses in the fourth generation.  The insensitivity of
the LO width $\Gamma_{\LO}$ with respect to the mass scale in the
fourth generation is reflecting the screening property of the
heavy-mass limit in this order.  The values of $\delta_{\myEW}^{(4)}$
are given for completeness but one should remember that the prediction
is in terms of ${\overline\Gamma}_{\LO}$.
\begin{table}
\begin{center}
\caption[]{\label{tab:EWNLOGG_NS}{
NLO EW corrections to the $\PH  \to \PGg\PGg$ decay width according to 
\eqn{bGamma} and estimate for the missing higher-order ($\delta_{_\THU}
$) corrections from \eqn{cTHU}.
Here we have fixed $\Mtp = 500\GeV, \Mlp=450\GeV$, and $\MH = 120\GeV$. 
}}
\vspace{1em}
\begin{tabular}{|c|c|cccc|}
\hline
\rule[-1ex]{0ex}{3.5ex}%
$\Mbp\,$[GeV] &
$\Mnp\,$[GeV] &
$\Gamma_{\LO}\, [\mathrm{GeV}]$ &
$\bar{\delta}_{\myEW}^{(4)}\,[\%]$  &
${\overline\Gamma}_{\LO}\, [\mathrm{GeV}]$ &
$\delta_{_\THU}\,[\%]$          \\
\hline
\rule[-0ex]{0ex}{2.5ex}%
$450$ & $350$ & $1.4656\cdot10^{-6}$ & $-96.1$ & $0.0576\cdot10^{-6}$ & $ 23.1$ \\ 
$450$ & $375$ & $1.4656\cdot10^{-6}$ & $-96.3$ & $0.0542\cdot10^{-6}$ & $ 23.8$ \\ 
$450$ & $400$ & $1.4656\cdot10^{-6}$ & $-96.5$ & $0.0507\cdot10^{-6}$ & $ 24.6$ \\ 
\hline                                                  
$500$ & $350$ & $1.4659\cdot10^{-6}$ & $-98.2$ & $0.0270\cdot10^{-6}$ & $ 33.8$ \\ 
$500$ & $375$ & $1.4659\cdot10^{-6}$ & $-98.3$ & $0.0247\cdot10^{-6}$ & $ 35.3$ \\ 
$500$ & $400$ & $1.4659\cdot10^{-6}$ & $-98.5$ & $0.0223\cdot10^{-6}$ & $ 37.1$ \\ 
\hline                                                  
$550$ & $350$ & $1.4662\cdot10^{-6}$ & $-99.5$ & $0.0067\cdot10^{-6}$ & $ 99.2$ \\
$550$ & $375$ & $1.4662\cdot10^{-6}$ & $-99.6$ & $0.0056\cdot10^{-6}$ & $>100$ \\
$550$ & $400$ & $1.4662\cdot10^{-6}$ & $-99.7$ & $0.0045\cdot10^{-6}$ & $>100$
\\
\hline
\end{tabular}
\end{center}
\end{table}
In the mass scenario \refeq{eq:scenarios_c}, the uncertainty from higher orders
$\delta_{_\THU}$ is large for low values of $\MH$.  In the extreme
scenario of \eqn{eq:masses}, above $\MH = 145\GeV$ the credibility of
our estimate for the effect of the NNLO corrections becomes more and
more questionable and the results cannot be trusted anymore, missing
the complete NNLO term.  In any case perturbation theory becomes
questionable for higher values of $\MH$.

It is worth noting that for $\PH \to VV$ (see \refse{se:H4f})
the situation is different.
There is no accidentally small LO (there SM3=SM4 in LO) and the square
of $A_{\NLO}$ is taken into account by the leading NNLO term taken
from \Bref{Djouadi:1997rj}, 
which serves as
our error estimate.

The decay mode $\PH\to \PGg\PZ$ is treated at LO only, since the NLO QCD
corrections within the SM3 are known to be small \cite{Spira:1991tj} and
can thus safely be neglected. The EW corrections in SM3 as
well as in SM4 are unknown. This implies a theoretical uncertainty of
the order of 100\% in the intermediate Higgs-boson mass range within SM4,
since large cancellations between the $\PW$ and fermion loops emerge at
LO similar to the decay mode $\PH\to \PGg\PGg$.

\section{Numerical results}
\label{sec:numres}

The results for the Higgs-boson production cross section via gluon
fusion have been obtained by including the NLO QCD corrections with
full quark-mass dependence \cite{Spira:1995rr} and the NNLO QCD
corrections in the limit of heavy quarks \cite{Anastasiou:2011qw}.
The full EW corrections \cite{Passarino:2011kv} have been included in
factorized form as discussed in \refse{sec:Hprod}. We use the
MSTW2008NNLO parton density functions \cite{Martin:2009iq} with the
strong coupling normalized to $\alphas(\MZ)=0.11707$ at NNLO.  The
renormalization and factorization scales are chosen as
$\mu_\mathrm{R}=\mu_\mathrm{F}=\MH/2$.

In \refta {tab:ggf-sm4-grid} we show results for the scenarios defined in
\refeq{eq:scenarios_a}/\refeq{eq:scenarios_b} for the Higgs production
cross section at $\sqrt{s}=8 \TeV$.%
\begin{table}
\caption{SM4 Higgs-boson production cross section via gluon fusion
including NNLO QCD and NLO EW corrections using MSTW2008NNLO PDFs for
$\sqrt{s}=8 \TeV$ in the scenarios
\refeq{eq:scenarios_a}/\refeq{eq:scenarios_b}.}
\label{tab:ggf-sm4-grid}
\vspace{1em}
\centerline{\input{plots/ggf_grid_8tev2}}
\end{table}
\begin{figure}
\centerline{\includegraphics[height=9cm, bb= 50 220 560 620]{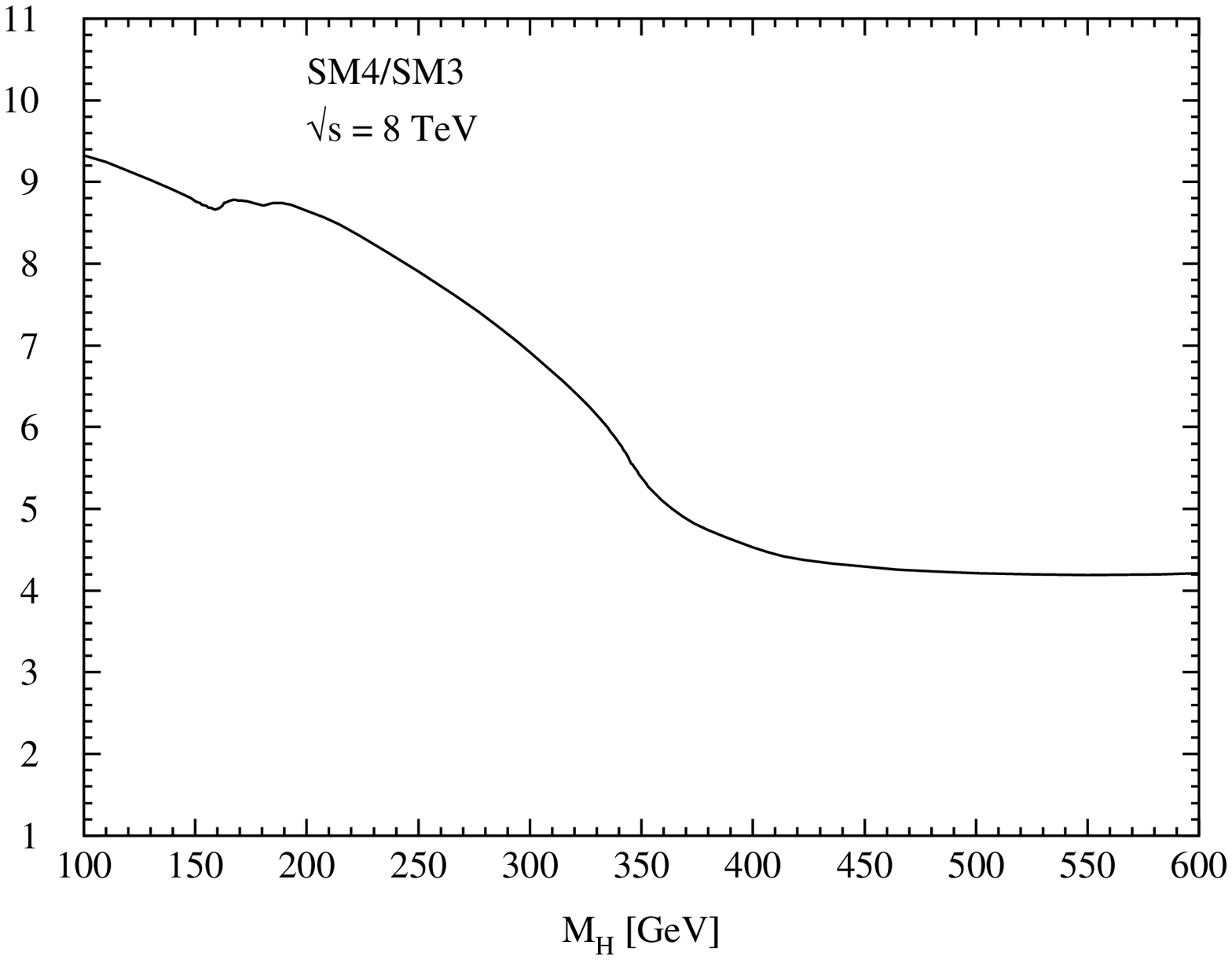}}
\caption{Ratio of Higgs-boson production cross sections via gluon
  fusion in SM4 with respect to SM3 including NNLO QCD and NLO EW
  corrections for $\Mtp = 500\GeV$, $\Mbp=450\GeV$, $\Mlp = 450\GeV$,
  and $\Mnp=375\GeV$ and $\sqrt{s}=8\TeV$.}
\label{fig:xs-sm4-ratio-8tev}
\end{figure}
For the specific scenario \refeq{eq:scenarios_c} we display the ratio
between the SM4 and SM3 cross sections at $8\TeV$ in
\reffi{fig:xs-sm4-ratio-8tev}.  The SM4 cross sections are enhanced by
factors of $4{-}9$ with respect to SM3.
In the extreme scenario \refeq{eq:masses} we have studied the gluon-fusion
cross section at $\sqrt{s}=7 \TeV$. Corresponding results are shown in
\refta{tab:ggf-sm4} and the ratio to the SM cross section is plotted
in \reffi{fig:xs-sm4-ratio}. The enhancement is similar as in the
scenario shown in \reffi{fig:xs-sm4-ratio-8tev}.
For the $\Pg\Pg$-fusion cross section in SM4 the QCD uncertainties are
about the same as in the SM3 case, while the additional uncertainties
due to the EW corrections have been discussed in \refse{sec:Hprod}.
\begin{table}
\caption{SM4 Higgs-boson production cross section via gluon fusion
including NNLO QCD and NLO EW corrections using MSTW2008NNLO PDFs for
$\sqrt{s}=7 \TeV$ in the extreme scenario \refeq{eq:masses}.}
\label{tab:ggf-sm4}
\vspace{1em}
\centerline{\input{plots/ggf}}
\end{table}
\begin{figure}
\centerline{\includegraphics[height=9cm, bb= 50 220 560 620]{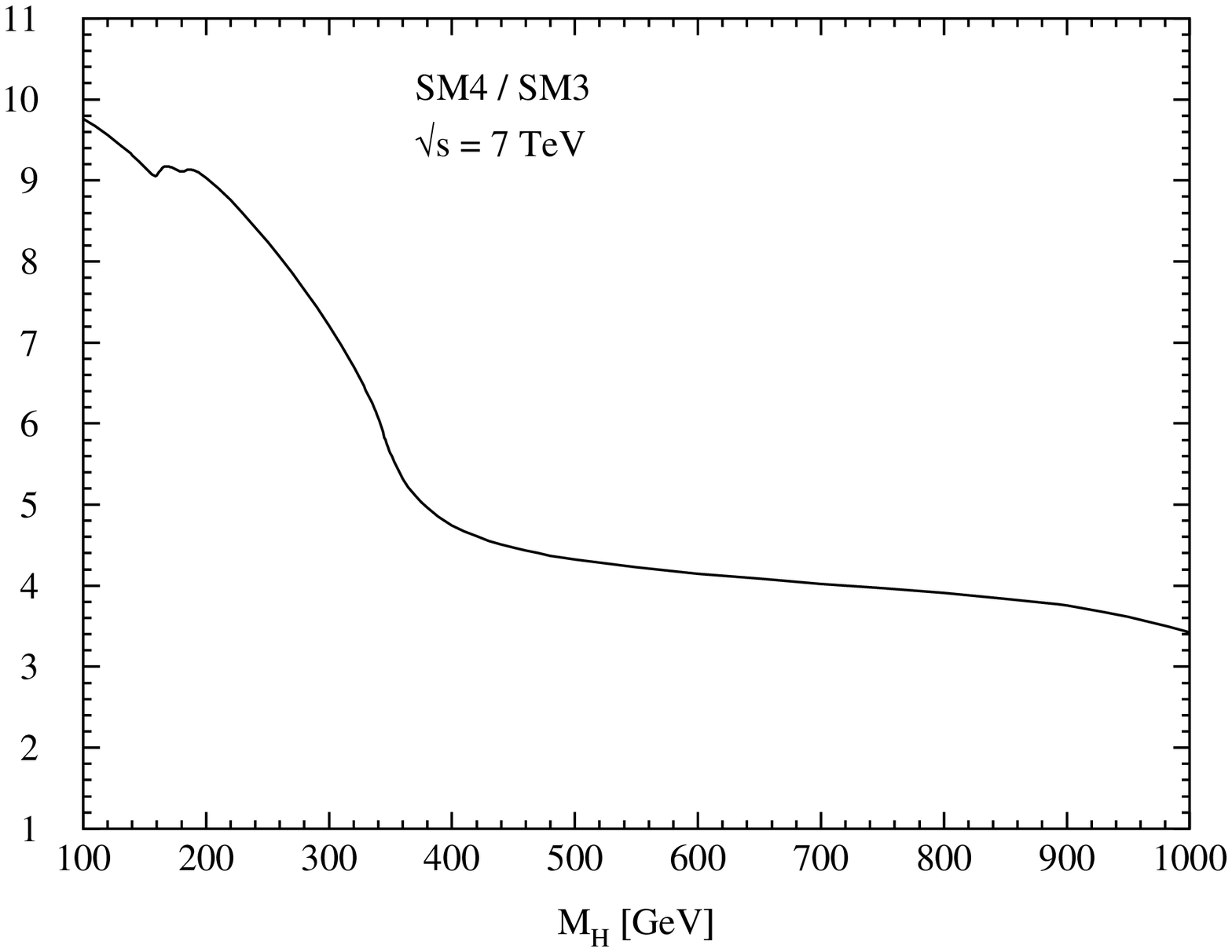}}
\caption{Ratio of Higgs-boson production cross sections via gluon
  fusion in SM4 with respect to SM3 including NNLO QCD and NLO EW
  corrections for $\sqrt{s}=7\TeV$ in the extreme scenario
  \refeq{eq:masses}.}
\label{fig:xs-sm4-ratio}
\end{figure}

The results for the Higgs branching fractions have been obtained in a
similar way as those for the results in SM3 in
\Brefs{Dittmaier:2011ti,Denner:2011mq}. While the partial widths for
$\PH\to\PW\PW/\PZ\PZ$ have been computed with {\sc Prophecy4f}, all
other partial widths have been calculated with {\sc
  HDECAY}.
Then, the branching ratios and the total width have been calculated
from these partial widths.

\begin{figure}            
\centerline{\includegraphics[height=9cm, bb= 155 560 385 785]{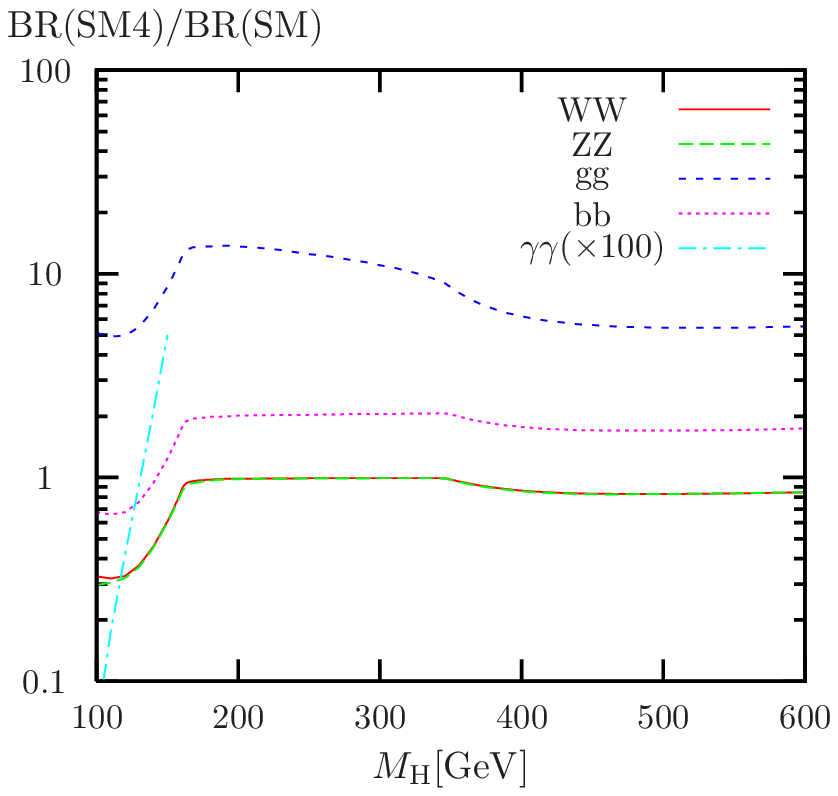}}
\caption{Ratio of branching fractions in SM4 with respect to SM3 for $\PW\PW$, $\PZ\PZ$, $\Pg\Pg$,
  $\PQb\bar{\PQb}$, and $\PGg\PGg$ decay channels ($\PGg\PGg$ ratio
  multiplied with $100$) as a function of $\MH$ for 
scenario \refeq{eq:scenarios_c}.}
\label{fig:br-sm4-ratio-grid}
\end{figure}

The results of the Higgs branching fractions for the scenarios defined
in \refeq{eq:scenarios_a}/\refeq{eq:scenarios_b} are shown in
\refta{tab:br-sm4-ff-grid} for the 2-fermion final states and in
\refta{tab:br-sm4-vv-grid} for the 2-gauge-boson final states.  In the
latter table also the total Higgs width is given.
Table~\ref{tab:br-sm4-comb-grid} lists the branching fractions for the
$\Pep\Pem\Pep\Pem$ and $\Pep\Pem\PGmp\PGmm$ final states as well as
several combined channels.  Apart from the sum of all 4-fermion final
states ($\PH \to 4\Pf$) the results for all-leptonic final states $\PH
\to 4\Pl$ with $\Pl = \Pe,\PGm,\PGt,\PGne,\PGnGm,\PGnGt$, the results
for all-hadronic final states $\PH \to 4\Pq$ with $\Pq =
\PQu,\PQd,\PQc,\PQs,\PQb$ and the semi-leptonic final states $\PH \to
2\Pl2\Pq$ are shown. To compare with the pure SM3,
\reffi{fig:br-sm4-ratio-grid} shows the ratios between the SM4 and SM3
branching fractions for the most important channels for 
the scenario \refeq{eq:scenarios_c}.
\begin{table}
\caption{SM4 Higgs branching fractions for 2-fermion decay channels
for the scenarios defined in \refeq{eq:scenarios_a}/\refeq{eq:scenarios_b}.} 
\label{tab:br-sm4-ff-grid}
\vspace{1em}
\centerline{\input{plots/br_ff_grid}}
\end{table}%
\begin{table}
\caption{SM4 Higgs branching fractions for 2-gauge-boson decay
  channels and total Higgs width for the scenarios defined in
  \refeq{eq:scenarios_a}/\refeq{eq:scenarios_b}.}
\label{tab:br-sm4-vv-grid}
\vspace{1em}
\centerline{\input{plots/br_vv_grid}}
\end{table}%
\begin{table}
\caption{SM4 Higgs branching fractions for 4-fermion final states
  with $\Pl = \Pe,\PGm,\PGt,\PGne,\PGnGm,\,\PGnGt$ and $\Pq = \PQu,\PQd,\PQc,\PQs,\PQb$ for the scenarios defined in
  \refeq{eq:scenarios_a}/\refeq{eq:scenarios_b}.}
\label{tab:br-sm4-comb-grid}
\vspace{1em}
\centerline{\input{plots/br_h4f_comb_grid}}
\end{table}%
While the branching ratio into gluons is enhanced by a factor
$5{-}15$, $\br(\PH\to\PQb\bar\PQb)$ is reduced for small $\MH$ but
enhanced for $\MH\gsim150\GeV$. The branching ratios into electroweak
gauge-boson pairs are suppressed for small Higgs masses, and the one
into photon pairs is reduced by $65$ to $100\%$ in the Higgs-mass
range $100\GeV<\MH<150\GeV$.

\begin{figure}
\centerline{\includegraphics[height=9cm, bb= 155 560 385 785]{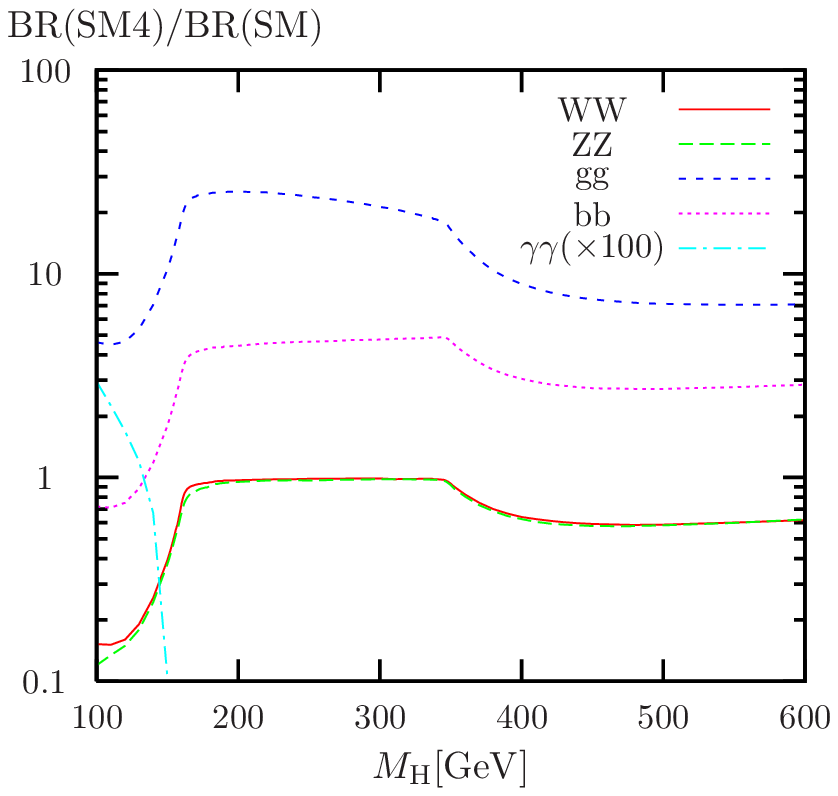}}
\caption{Ratio of branching fractions in SM4 with respect to SM3 for $\PW\PW$, $\PZ\PZ$, $\Pg\Pg$,
  $\PQb\bar{\PQb}$, and $\PGg\PGg$ decay channels ($\PGg\PGg$ ratio
  multiplied with $100$) as a function of $\MH$ in the extreme
  scenario \refeq{eq:masses}.}
\label{fig:br-sm4-ratio}
\end{figure}

Results in the extreme scenario \refeq{eq:masses} for Higgs masses up
to $1 \TeV$ are shown in \refta{tab:br-sm4-ff} for the 2-fermion final
states, in \refta{tab:br-sm4-vv} for the 2-gauge-boson final states
and the total Higgs width, and in Table~\ref{tab:br-sm4-comb} for
selected 4-fermion final states.  The ratios between the SM4 and SM3
branching fractions for the most important channels are shown in
\reffi{fig:br-sm4-ratio}. As compared to the scenario of
\reffi{fig:br-sm4-ratio-grid}, the enhancement and suppression effects
are stronger (as they scale roughly with the square of the heavy
fermion masses). While $\br(\PH\to\gamma\gamma)$ is different in
detail it is again suppressed by a factor 100.
\begin{table}
\caption{SM4 Higgs branching fractions for 2-fermion decay channels
in the extreme scenario \refeq{eq:masses}.}
\label{tab:br-sm4-ff}
\vspace{1em}
\centerline{\input{plots/br_ff}}
\end{table}
\begin{table}
\caption{SM4 Higgs branching fractions for 2-gauge-boson decay
  channels and total Higgs width in the extreme scenario \refeq{eq:masses}.}
\label{tab:br-sm4-vv}
\vspace{1em}
\centerline{\input{plots/br_vv}}
\end{table}
\begin{table}
\caption{SM4 Higgs branching fractions for 4-fermion final states
  with $\Pl = \Pe,\PGm,\PGt,\PGne,\PGnGm,\,\PGnGt$ and $\Pq =
  \PQu,\PQd,\PQc,\PQs,\PQb$ in the extreme scenario \refeq{eq:masses}.}
\label{tab:br-sm4-comb}
\vspace{1em}
\centerline{\input{plots/br_h4f_comb}}
\end{table}

The effect of the NLO EW corrections on the $\PH \to \PGg\PGg$ decay
width in the scenarios \refeq{eq:scenarios_a}/\refeq{eq:scenarios_b}
is shown in \refta{tab:EWNLOBRGG_grid}.  The branching ratio for $\PH
\to \PGg\PGg$ is strongly reduced in SM4 owing to cancellations
between LO and NLO.
\begin{table}
\begin{center}
\caption[]{\label{tab:EWNLOBRGG_grid}{Higgs branching fractions for
    the $\PGg\PGg$ decay channel without and with NLO EW corrections
    in the scenarios \refeq{eq:scenarios_a}/\refeq{eq:scenarios_b}
    (QCD corrections are always included).}}
\vspace{1em} \input{plots/BR_gaga_scan_grid}
\end{center}
\end{table}
In \refta{tab:EWNLOBRGG} we display the effect of the NLO EW
corrections on the $\PH \to \PGg\PGg$ decay width in the extreme
scenario \refeq{eq:masses}. While the branching ratio differs
considerably from those in scenarios
\refeq{eq:scenarios_a}/\refeq{eq:scenarios_b} a similarly strong
reduction by a factor of 100 with respect to SM3 is observed. Thus,
this branching ratio is completely irrelevant in SM4.

\clearpage

\begin{table}    
\begin{center}
\caption[]{\label{tab:EWNLOBRGG}{Higgs branching fractions for the
    $\PGg\PGg$ decay channel without and with NLO EW corrections in
    the extreme scenario \refeq{eq:masses} (QCD corrections are always
    included).}} \vspace{1em}
\begin{tabular}{|ccc|}
\hline
\rule[-1ex]{0ex}{3.5ex}%
$\MH\,$[GeV] &
w/o NLO EW       &
w/ NLO EW           \\
\hline
\rule[0ex]{0ex}{2.5ex}%
$100$ & $1.31\,\cdot\,10^{-4}$  & $4.65\,\cdot\,10^{-5}$ \\
$110$ & $1.72\,\cdot\,10^{-4}$  & $4.40\,\cdot\,10^{-5}$ \\
$120$ & $2.26\,\cdot\,10^{-4}$  & $3.77\,\cdot\,10^{-5}$ \\
$130$ & $2.95\,\cdot\,10^{-4}$  & $2.71\,\cdot\,10^{-5}$ \\
$140$ & $3.81\,\cdot\,10^{-4}$  & $1.30\,\cdot\,10^{-5}$ \\
$150$ & $4.74\,\cdot\,10^{-4}$  & $1.42\,\cdot\,10^{-6}$%
\rule[-1ex]{0ex}{2.5ex}\\
\hline
\end{tabular}
\end{center}
\end{table}

\section{Conclusions}
\label{sec:concl}

Additional hypothetical heavy-fermion generations, which are embedded
in the Standard Model, strongly affect the prediction for the
production and decay of a Higgs boson.  The Yukawa couplings of heavy
fermions grow very large, eventually jeopardizing the use of
perturbation theory.

In this article we have presented state-of-the-art predictions for the
Higgs-boson production cross section via gluon fusion and for all
relevant Higgs-boson decay channels including one additional
heavy-fermion generation in a variety of scenarios with a
generic mass scale of $450\GeV$ as well as for an extreme scenario with
a mass scale of $600\GeV$, which is at the border between perturbativity 
and non-perturbativity in the 4th-generation sector.  
The loop-induced transitions $\Pg\Pg\to\PH$, $\PH\to\Pg\Pg$,
$\PH\to\gamma\gamma$ receive large lowest-order contributions, as
frequently pointed out in the literature before.  Here we emphasize
the effect that on top of that the electroweak radiative corrections
grow very large. They typically grow with powers of the heavy-fermion
masses, eventually leading to a breakdown of perturbation theory.
For Higgs production via gluon fusion and the Higgs decay into gluon
pairs they are at the level of 10\% for $\MH<600\GeV$.  For the
important Higgs decays into WW or ZZ pairs we find corrections of the
order of $-40\%$ and $-60\%$ or more for the adopted heavy-fermion
mass scales of $450\GeV$ and $600\GeV$, respectively, where the onset
of the non-perturbative regime is clearly visible by electroweak
one-loop corrections of the size of about $-85\%$ in the latter case.
The branching ratios into fermion pairs are enhanced by $30\%$ and
$60\%$ for 4th-generation fermion masses of about $450\GeV$ and
$600\GeV$, respectively.  The branching ratio for the decay into
photon pairs is reduced by $65$ to $100\%$ in the Higgs-mass range
$100\GeV<\MH<150\GeV$ in all considered scenarios for a heavy 4th
fermion generation, where the reduction factor, however, shows a
strong dependence on the Higgs and heavy-fermion masses.  We also
present estimates for the respective theoretical uncertainties, which
are quite large (several $10\%$).  As the NLO EW corrections are
enhanced by powers of the masses of the heavy fermions they depend
strongly on the actual values of these masses.

The presented results and error estimates, the qualitative description of the
most important impact of heavy fermions, and the description of the
available tools and calculations will certainly prove useful in upcoming
refined analyses of LHC data on Higgs searches.

\subsection*{Acknowledgements}

This work is supported in part by the Gottfried Wilhelm
Leibniz programme of the Deutsche Forschungsgemeinschaft (DFG)
and by Ministero dell'Istruzione, dell'Universit\`a e della Ricerca
(MIUR) under contract 2008H8F9RA$\_$002.  We gratefully acknowledge
several discussions with P.~Gambino, C.~Mariotti, and R.~Tanaka.

\end{document}

%% file: plots/SM4EW_ctab.tex
\tabcolsep 4pt
\begin{tabular}{|cc@{\hspace{6pt}}|cc@{\hspace{6pt}}|}
\hline
\rule[-1ex]{0ex}{3.5ex}%
$\MH$ [GeV] & $\delta^{(4)}_{\rm EW}$ [\%] &
$\MH$ [GeV] & $\delta^{(4)}_{\rm EW}$ [\%] \\
\hline
\rule[-0ex]{0ex}{2.5ex}%
$100$ & $7.08$ & $180$ & $\phantom{+}3.22$ \\
$110$ & $7.01$ & $190$ & $\phantom{+}2.79$ \\
$120$ & $6.91$ & $200$ & $\phantom{+}2.20$ \\
$130$ & $6.77$ & $250$ & $\phantom{+}0.39$ \\
$140$ & $6.55$ & $300$ & $         - 1.11$ \\
$150$ & $6.16$ & $400$ & $         - 3.84$ \\
$160$ & $4.87$ & $500$ & $         - 8.71$ \\
$170$ & $4.38$ & $600$ & $         - 17.00$
\rule[-1ex]{0ex}{2.5ex}\\
\hline
\end{tabular}

%% file: plots/SM4EW_ctabscan.tex
\tabcolsep 4pt
\begin{tabular}{|cc|c||cc|c||cc|c|}
\hline
\multicolumn{3}{|c||}{ $\MH=120 \GeV$ }&
\multicolumn{3}{|c||}{ $\MH=350 \GeV$ }&
\multicolumn{3}{|c|}{ $\MH=600 \GeV$ }\\
\hline
\rule[-1ex]{0ex}{3.5ex}%
$\Mbp$&
$\Mnp$&
$\delta^{(4)}_{\rm EW}$[\%]& 
$\Mbp$&
$\Mnp$&
$\delta^{(4)}_{\rm EW}$[\%]& 
$\Mbp$&
$\Mnp$&
$\delta^{(4)}_{\rm EW}$[\%]\\
\multicolumn{2}{|c|}{ in $\GeV$ }& &
\multicolumn{2}{|c|}{ in $\GeV$ }& &
\multicolumn{2}{|c|}{ in $\GeV$ }& \\
\hline
\rule[-0ex]{0ex}{2.5ex}%
$450$ &$350$ &$6.72$ &$450$ &$350$ &$-4.25$ & $450$ &$300$ &$-20.27$\\
$450$ &$375$ &$6.91$ &$450$ &$375$ &$-4.05$ & $450$ &$350$ &$-17.41$\\
$450$ &$400$ &$7.14$ &$450$ &$400$ &$-3.82$ & $450$ &$400$ &$-16.63$\\
$500$ &$350$ &$6.61$ &$500$ &$350$ &$-4.21$ & $500$ &$300$ &$-20.67$\\
$500$ &$375$ &$6.81$ &$500$ &$375$ &$-4.01$ & $500$ &$350$ &$-17.80$\\
$500$ &$400$ &$7.03$ &$500$ &$400$ &$-3.78$ & $500$ &$400$ &$-17.03$\\
$550$ &$350$ &$6.72$ &$550$ &$350$ &$-3.93$ & $550$ &$300$ &$-20.82$\\
$550$ &$375$ &$6.91$ &$550$ &$375$ &$-3.73$ & $550$ &$350$ &$-17.95$\\
$550$ &$400$ &$7.14$ &$550$ &$400$ &$-3.50$ & $550$ &$400$ &$-17.17$\\
\hline
\end{tabular}

%% file: plots/ggf_grid_8tev2.tex
\tabcolsep 4pt
\begin{tabular}{|cccc|cccc|}
\hline
\rule[-1ex]{0ex}{3.5ex}%
$\MH$ [GeV] & $\Mbp$ [GeV] & $\Mnp$ [GeV] & $\sigma$ [pb] &
$\MH$ [GeV] & $\Mbp$ [GeV] & $\Mnp$ [GeV] & $\sigma$ [pb] \\
\hline
\rule[-0ex]{0ex}{2.5ex}%
 120  &   450  &   350  &   199.6  &   350  &   500  &   400  &   9.946  \\
 120  &   450  &   375  &   199.9  &   350  &   550  &   350  &   9.899  \\
 120  &   450  &   400  &   200.3  &   350  &   550  &   375  &   9.920  \\
 120  &   500  &   350  &   199.2  &   350  &   550  &   400  &   9.944  \\

 120  &   500  &   375  &   199.6  &   600  &   450  &   300  &   1.236  \\
 120  &   500  &   400  &   200.0  &   600  &   450  &   350  &   1.280 \\  
 120  &   550  &   350  &   199.3  &   600  &   450  &   400  &   1.292  \\
 120  &   550  &   375  &   199.7  &   600  &   500  &   300  &   1.209  \\
 120  &   550  &   400  &   200.1  &   600  &   500  &   350  &   1.253  \\

 350  &   450  &   350  &   9.940  &   600  &   500  &   400  &   1.271  \\
 350  &   450  &   375  &   9.961  &   600  &   550  &   300  &   1.193 \\
 350  &   450  &   400  &   9.986  &   600  &   550  &   350  &   1.236 \\
 350  &   500  &   350  &   9.901  &   600  &   550  &   400  &   1.248  \\
 350  &   500  &   375  &   9.922  &        &        &        &
\rule[-1ex]{0ex}{2.5ex}\\
\hline
\end{tabular}

%% file: plots/ggf.tex
\tabcolsep 4pt
\begin{tabular}{|cc|cc|cc|cc|cc|}
\hline
\rule[-1ex]{0ex}{3.5ex}%
$\MH$ [GeV] & $\sigma$ [pb] &
$\MH$ [GeV] & $\sigma$ [pb] \\
\hline
\rule[-0ex]{0ex}{2.5ex}%
100 & 244  & 200 & 48.6    \\ 
110 & 199  & 250 & 27.7    \\
120 & 165  & 300 & 17.6    \\ 
130 & 138  & 400 & 9.59    \\
140 & 117  & 500  & 3.70   \\
150 & 99.2 & 600  & 1.40   \\
160 & 84.5 & 700  & 0.556  \\
170 & 73.0 & 800  & 0.235  \\
180 & 63.1 & 900  & 0.104  \\
190 & 55.2 & 1000 & 0.0456%
\rule[-1ex]{0ex}{2.5ex}\\
\hline
\end{tabular}

%% file: plots/br_ff_grid.tex
\begin{tabular}{|ccccccc|}
\hline $\vphantom{\bar{\bar{\bar{\PQb}}}}$%
$\MH$ / \Mbp / \Mnp &   $\Hbb$ & $\Htautau$ & $\Hmumu$ &   $\Hss$ &   $\Hcc$ &   $\Htt$ \\ 
{}[GeV]& & & & & & \\\hline
  120/  450/  350 & $4.39\cdot 10^{-1}$ & $4.77\cdot 10^{-2}$ & $1.65\cdot 10^{-4}$ & $1.87\cdot 10^{-4}$ & $2.21\cdot 10^{-2}$ & $0.00$ \\
  120/  450/  375 & $4.39\cdot 10^{-1}$ & $4.77\cdot 10^{-2}$ & $1.66\cdot 10^{-4}$ & $1.87\cdot 10^{-4}$ & $2.22\cdot 10^{-2}$ & $0.00$ \\
  120/  450/  400 & $4.39\cdot 10^{-1}$ & $4.77\cdot 10^{-2}$ & $1.66\cdot 10^{-4}$ & $1.87\cdot 10^{-4}$ & $2.22\cdot 10^{-2}$ & $0.00$ \\
  120/  500/  350 & $4.45\cdot 10^{-1}$ & $4.83\cdot 10^{-2}$ & $1.68\cdot 10^{-4}$ & $1.90\cdot 10^{-4}$ & $2.24\cdot 10^{-2}$ & $0.00$ \\
  120/  500/  375 & $4.45\cdot 10^{-1}$ & $4.84\cdot 10^{-2}$ & $1.68\cdot 10^{-4}$ & $1.90\cdot 10^{-4}$ & $2.25\cdot 10^{-2}$ & $0.00$ \\
  120/  500/  400 & $4.45\cdot 10^{-1}$ & $4.84\cdot 10^{-2}$ & $1.68\cdot 10^{-4}$ & $1.90\cdot 10^{-4}$ & $2.25\cdot 10^{-2}$ & $0.00$ \\
  120/  550/  350 & $4.52\cdot 10^{-1}$ & $4.91\cdot 10^{-2}$ & $1.70\cdot 10^{-4}$ & $1.93\cdot 10^{-4}$ & $2.28\cdot 10^{-2}$ & $0.00$ \\
  120/  550/  375 & $4.52\cdot 10^{-1}$ & $4.91\cdot 10^{-2}$ & $1.70\cdot 10^{-4}$ & $1.93\cdot 10^{-4}$ & $2.28\cdot 10^{-2}$ & $0.00$ \\
  120/  550/  400 & $4.52\cdot 10^{-1}$ & $4.92\cdot 10^{-2}$ & $1.71\cdot 10^{-4}$ & $1.93\cdot 10^{-4}$ & $2.28\cdot 10^{-2}$ & $0.00$ \\
  350/  450/  350 & $7.25\cdot 10^{-4}$ & $9.60\cdot 10^{-5}$ & $3.33\cdot 10^{-7}$ & $3.09\cdot 10^{-7}$ & $3.64\cdot 10^{-5}$ & $3.14\cdot 10^{-2}$ \\
  350/  450/  375 & $7.32\cdot 10^{-4}$ & $9.68\cdot 10^{-5}$ & $3.36\cdot 10^{-7}$ & $3.12\cdot 10^{-7}$ & $3.68\cdot 10^{-5}$ & $3.17\cdot 10^{-2}$ \\
  350/  450/  400 & $7.39\cdot 10^{-4}$ & $9.78\cdot 10^{-5}$ & $3.39\cdot 10^{-7}$ & $3.15\cdot 10^{-7}$ & $3.71\cdot 10^{-5}$ & $3.20\cdot 10^{-2}$ \\
  350/  500/  350 & $7.72\cdot 10^{-4}$ & $1.02\cdot 10^{-4}$ & $3.54\cdot 10^{-7}$ & $3.29\cdot 10^{-7}$ & $3.88\cdot 10^{-5}$ & $3.35\cdot 10^{-2}$ \\
  350/  500/  375 & $7.79\cdot 10^{-4}$ & $1.03\cdot 10^{-4}$ & $3.57\cdot 10^{-7}$ & $3.32\cdot 10^{-7}$ & $3.92\cdot 10^{-5}$ & $3.38\cdot 10^{-2}$ \\
  350/  500/  400 & $7.87\cdot 10^{-4}$ & $1.04\cdot 10^{-4}$ & $3.61\cdot 10^{-7}$ & $3.35\cdot 10^{-7}$ & $3.95\cdot 10^{-5}$ & $3.41\cdot 10^{-2}$ \\
  350/  550/  350 & $8.36\cdot 10^{-4}$ & $1.11\cdot 10^{-4}$ & $3.83\cdot 10^{-7}$ & $3.56\cdot 10^{-7}$ & $4.20\cdot 10^{-5}$ & $3.62\cdot 10^{-2}$ \\
  350/  550/  375 & $8.44\cdot 10^{-4}$ & $1.12\cdot 10^{-4}$ & $3.87\cdot 10^{-7}$ & $3.59\cdot 10^{-7}$ & $4.24\cdot 10^{-5}$ & $3.66\cdot 10^{-2}$ \\
  350/  550/  400 & $8.53\cdot 10^{-4}$ & $1.13\cdot 10^{-4}$ & $3.91\cdot 10^{-7}$ & $3.63\cdot 10^{-7}$ & $4.28\cdot 10^{-5}$ & $3.70\cdot 10^{-2}$ \\
  600/  450/  300 & $1.24\cdot 10^{-4}$ & $1.80\cdot 10^{-5}$ & $6.25\cdot 10^{-8}$ & $5.26\cdot 10^{-8}$ & $6.20\cdot 10^{-6}$ & $2.97\cdot 10^{-1}$ \\
  600/  450/  350 & $1.22\cdot 10^{-4}$ & $1.78\cdot 10^{-5}$ & $6.17\cdot 10^{-8}$ & $5.19\cdot 10^{-8}$ & $6.12\cdot 10^{-6}$ & $2.93\cdot 10^{-1}$ \\
  600/  450/  400 & $1.23\cdot 10^{-4}$ & $1.80\cdot 10^{-5}$ & $6.24\cdot 10^{-8}$ & $5.24\cdot 10^{-8}$ & $6.18\cdot 10^{-6}$ & $2.96\cdot 10^{-1}$ \\
  600/  500/  300 & $1.29\cdot 10^{-4}$ & $1.88\cdot 10^{-5}$ & $6.51\cdot 10^{-8}$ & $5.49\cdot 10^{-8}$ & $6.47\cdot 10^{-6}$ & $3.10\cdot 10^{-1}$ \\
  600/  500/  350 & $1.27\cdot 10^{-4}$ & $1.86\cdot 10^{-5}$ & $6.43\cdot 10^{-8}$ & $5.42\cdot 10^{-8}$ & $6.39\cdot 10^{-6}$ & $3.06\cdot 10^{-1}$ \\
  600/  500/  400 & $1.29\cdot 10^{-4}$ & $1.88\cdot 10^{-5}$ & $6.50\cdot 10^{-8}$ & $5.48\cdot 10^{-8}$ & $6.46\cdot 10^{-6}$ & $3.09\cdot 10^{-1}$ \\
  600/  550/  300 & $1.36\cdot 10^{-4}$ & $1.98\cdot 10^{-5}$ & $6.87\cdot 10^{-8}$ & $5.80\cdot 10^{-8}$ & $6.84\cdot 10^{-6}$ & $3.27\cdot 10^{-1}$ \\
  600/  550/  350 & $1.35\cdot 10^{-4}$ & $1.96\cdot 10^{-5}$ & $6.78\cdot 10^{-8}$ & $5.72\cdot 10^{-8}$ & $6.75\cdot 10^{-6}$ & $3.23\cdot 10^{-1}$ \\
  600/  550/  400 & $1.36\cdot 10^{-4}$ & $1.98\cdot 10^{-5}$ & $6.85\cdot 10^{-8}$ & $5.78\cdot 10^{-8}$ & $6.82\cdot 10^{-6}$ & $3.26\cdot 10^{-1}$ \\
\hline\end{tabular}

%% file: plots/br_vv_grid.tex
\begin{tabular}{|cccccc|}
\hline$\MH$ / \Mbp / \Mnp &   $\Hgg$ &  $\HZga$ &   $\HWW$ &   $\HZZ$ & $\Gamma_{\PH}$ \\ 
{}[GeV]& & & & & [GeV] \\\hline
  120/  450/  350 & $4.39\cdot 10^{-1}$ & $4.54\cdot 10^{-4}$ & $4.70\cdot 10^{-2}$ & $5.14\cdot 10^{-3}$ & $6.68\cdot 10^{-3}$ \\
  120/  450/  375 & $4.39\cdot 10^{-1}$ & $4.54\cdot 10^{-4}$ & $4.66\cdot 10^{-2}$ & $5.10\cdot 10^{-3}$ & $6.69\cdot 10^{-3}$ \\
  120/  450/  400 & $4.39\cdot 10^{-1}$ & $4.53\cdot 10^{-4}$ & $4.62\cdot 10^{-2}$ & $5.05\cdot 10^{-3}$ & $6.70\cdot 10^{-3}$ \\
  120/  500/  350 & $4.34\cdot 10^{-1}$ & $4.51\cdot 10^{-4}$ & $4.47\cdot 10^{-2}$ & $4.88\cdot 10^{-3}$ & $6.74\cdot 10^{-3}$ \\
  120/  500/  375 & $4.35\cdot 10^{-1}$ & $4.50\cdot 10^{-4}$ & $4.43\cdot 10^{-2}$ & $4.84\cdot 10^{-3}$ & $6.75\cdot 10^{-3}$ \\
  120/  500/  400 & $4.35\cdot 10^{-1}$ & $4.50\cdot 10^{-4}$ & $4.38\cdot 10^{-2}$ & $4.79\cdot 10^{-3}$ & $6.76\cdot 10^{-3}$ \\
  120/  550/  350 & $4.29\cdot 10^{-1}$ & $4.45\cdot 10^{-4}$ & $4.19\cdot 10^{-2}$ & $4.55\cdot 10^{-3}$ & $6.82\cdot 10^{-3}$ \\
  120/  550/  375 & $4.29\cdot 10^{-1}$ & $4.45\cdot 10^{-4}$ & $4.15\cdot 10^{-2}$ & $4.51\cdot 10^{-3}$ & $6.83\cdot 10^{-3}$ \\
  120/  550/  400 & $4.30\cdot 10^{-1}$ & $4.44\cdot 10^{-4}$ & $4.10\cdot 10^{-2}$ & $4.46\cdot 10^{-3}$ & $6.84\cdot 10^{-3}$ \\
  350/  450/  350 & $6.91\cdot 10^{-3}$ & $5.54\cdot 10^{-5}$ & $6.62\cdot 10^{-1}$ & $2.99\cdot 10^{-1}$ & $9.72$ \\
  350/  450/  375 & $6.97\cdot 10^{-3}$ & $5.58\cdot 10^{-5}$ & $6.62\cdot 10^{-1}$ & $2.98\cdot 10^{-1}$ & $9.65$ \\
  350/  450/  400 & $7.04\cdot 10^{-3}$ & $5.62\cdot 10^{-5}$ & $6.61\cdot 10^{-1}$ & $2.98\cdot 10^{-1}$ & $9.58$ \\
  350/  500/  350 & $7.17\cdot 10^{-3}$ & $5.77\cdot 10^{-5}$ & $6.60\cdot 10^{-1}$ & $2.98\cdot 10^{-1}$ & $9.33$ \\
  350/  500/  375 & $7.24\cdot 10^{-3}$ & $5.81\cdot 10^{-5}$ & $6.60\cdot 10^{-1}$ & $2.98\cdot 10^{-1}$ & $9.26$ \\
  350/  500/  400 & $7.31\cdot 10^{-3}$ & $5.86\cdot 10^{-5}$ & $6.59\cdot 10^{-1}$ & $2.98\cdot 10^{-1}$ & $9.19$ \\
  350/  550/  350 & $7.54\cdot 10^{-3}$ & $6.07\cdot 10^{-5}$ & $6.58\cdot 10^{-1}$ & $2.97\cdot 10^{-1}$ & $8.87$ \\
  350/  550/  375 & $7.62\cdot 10^{-3}$ & $6.12\cdot 10^{-5}$ & $6.58\cdot 10^{-1}$ & $2.96\cdot 10^{-1}$ & $8.80$ \\
  350/  550/  400 & $7.70\cdot 10^{-3}$ & $6.17\cdot 10^{-5}$ & $6.58\cdot 10^{-1}$ & $2.96\cdot 10^{-1}$ & $8.73$ \\
  600/  450/  300 & $2.27\cdot 10^{-3}$ & $6.52\cdot 10^{-6}$ & $4.71\cdot 10^{-1}$ & $2.30\cdot 10^{-1}$ & $8.96\cdot 10^{1}$ \\
  600/  450/  350 & $2.32\cdot 10^{-3}$ & $6.43\cdot 10^{-6}$ & $4.73\cdot 10^{-1}$ & $2.31\cdot 10^{-1}$ & $9.09\cdot 10^{1}$ \\
  600/  450/  400 & $2.36\cdot 10^{-3}$ & $6.47\cdot 10^{-6}$ & $4.71\cdot 10^{-1}$ & $2.30\cdot 10^{-1}$ & $9.03\cdot 10^{1}$ \\
  600/  500/  300 & $2.27\cdot 10^{-3}$ & $6.62\cdot 10^{-6}$ & $4.62\cdot 10^{-1}$ & $2.26\cdot 10^{-1}$ & $8.78\cdot 10^{1}$ \\
  600/  500/  350 & $2.31\cdot 10^{-3}$ & $6.53\cdot 10^{-6}$ & $4.65\cdot 10^{-1}$ & $2.27\cdot 10^{-1}$ & $8.92\cdot 10^{1}$ \\
  600/  500/  400 & $2.35\cdot 10^{-3}$ & $6.57\cdot 10^{-6}$ & $4.63\cdot 10^{-1}$ & $2.26\cdot 10^{-1}$ & $8.86\cdot 10^{1}$ \\
  600/  550/  300 & $2.29\cdot 10^{-3}$ & $6.77\cdot 10^{-6}$ & $4.51\cdot 10^{-1}$ & $2.20\cdot 10^{-1}$ & $8.57\cdot 10^{1}$ \\
  600/  550/  350 & $2.34\cdot 10^{-3}$ & $6.67\cdot 10^{-6}$ & $4.54\cdot 10^{-1}$ & $2.21\cdot 10^{-1}$ & $8.70\cdot 10^{1}$ \\
  600/  550/  400 & $2.38\cdot 10^{-3}$ & $6.71\cdot 10^{-6}$ & $4.51\cdot 10^{-1}$ & $2.20\cdot 10^{-1}$ & $8.64\cdot 10^{1}$ \\
\hline\end{tabular}

%% file: plots/br_h4f_comb_grid.tex
\begin{tabular}{|ccccccc|}
\hline$\MH$ / \Mbp / \Mnp &   $\PH \to 4\Pe$ & $\PH \to 2\Pe 2\PGm$ &   $\PH \to 4\Pl$ &   $\PH \to 4\Pq$ & $\PH \to 2\Pl 2\Pq$ &   $\PH \to 4\Pf$ \\ 
{}[GeV]& & & & & & \\\hline
  120/  450/  350 & $6.39\cdot 10^{-6}$ & $1.14\cdot 10^{-5}$ & $5.24\cdot 10^{-3}$ & $2.38\cdot 10^{-2}$ & $2.27\cdot 10^{-2}$ & $5.17\cdot 10^{-2}$ \\
  120/  450/  375 & $6.33\cdot 10^{-6}$ & $1.13\cdot 10^{-5}$ & $5.19\cdot 10^{-3}$ & $2.36\cdot 10^{-2}$ & $2.25\cdot 10^{-2}$ & $5.12\cdot 10^{-2}$ \\
  120/  450/  400 & $6.27\cdot 10^{-6}$ & $1.12\cdot 10^{-5}$ & $5.14\cdot 10^{-3}$ & $2.33\cdot 10^{-2}$ & $2.23\cdot 10^{-2}$ & $5.08\cdot 10^{-2}$ \\
  120/  500/  350 & $6.05\cdot 10^{-6}$ & $1.08\cdot 10^{-5}$ & $4.96\cdot 10^{-3}$ & $2.26\cdot 10^{-2}$ & $2.15\cdot 10^{-2}$ & $4.91\cdot 10^{-2}$ \\
  120/  500/  375 & $5.99\cdot 10^{-6}$ & $1.07\cdot 10^{-5}$ & $4.91\cdot 10^{-3}$ & $2.24\cdot 10^{-2}$ & $2.13\cdot 10^{-2}$ & $4.87\cdot 10^{-2}$ \\
  120/  500/  400 & $5.93\cdot 10^{-6}$ & $1.06\cdot 10^{-5}$ & $4.86\cdot 10^{-3}$ & $2.22\cdot 10^{-2}$ & $2.11\cdot 10^{-2}$ & $4.82\cdot 10^{-2}$ \\
  120/  550/  350 & $5.62\cdot 10^{-6}$ & $1.00\cdot 10^{-5}$ & $4.63\cdot 10^{-3}$ & $2.12\cdot 10^{-2}$ & $2.02\cdot 10^{-2}$ & $4.60\cdot 10^{-2}$ \\
  120/  550/  375 & $5.56\cdot 10^{-6}$ & $9.93\cdot 10^{-6}$ & $4.58\cdot 10^{-3}$ & $2.10\cdot 10^{-2}$ & $2.00\cdot 10^{-2}$ & $4.56\cdot 10^{-2}$ \\
  120/  550/  400 & $5.50\cdot 10^{-6}$ & $9.82\cdot 10^{-6}$ & $4.53\cdot 10^{-3}$ & $2.08\cdot 10^{-2}$ & $1.98\cdot 10^{-2}$ & $4.51\cdot 10^{-2}$ \\
  350/  450/  350 & $3.25\cdot 10^{-4}$ & $6.52\cdot 10^{-4}$ & $9.47\cdot 10^{-2}$ & $4.52\cdot 10^{-1}$ & $4.14\cdot 10^{-1}$ & $9.61\cdot 10^{-1}$ \\
  350/  450/  375 & $3.25\cdot 10^{-4}$ & $6.52\cdot 10^{-4}$ & $9.46\cdot 10^{-2}$ & $4.52\cdot 10^{-1}$ & $4.14\cdot 10^{-1}$ & $9.60\cdot 10^{-1}$ \\
  350/  450/  400 & $3.24\cdot 10^{-4}$ & $6.51\cdot 10^{-4}$ & $9.45\cdot 10^{-2}$ & $4.52\cdot 10^{-1}$ & $4.13\cdot 10^{-1}$ & $9.60\cdot 10^{-1}$ \\
  350/  500/  350 & $3.23\cdot 10^{-4}$ & $6.48\cdot 10^{-4}$ & $9.41\cdot 10^{-2}$ & $4.52\cdot 10^{-1}$ & $4.12\cdot 10^{-1}$ & $9.58\cdot 10^{-1}$ \\
  350/  500/  375 & $3.23\cdot 10^{-4}$ & $6.48\cdot 10^{-4}$ & $9.40\cdot 10^{-2}$ & $4.52\cdot 10^{-1}$ & $4.12\cdot 10^{-1}$ & $9.58\cdot 10^{-1}$ \\
  350/  500/  400 & $3.22\cdot 10^{-4}$ & $6.47\cdot 10^{-4}$ & $9.39\cdot 10^{-2}$ & $4.52\cdot 10^{-1}$ & $4.12\cdot 10^{-1}$ & $9.58\cdot 10^{-1}$ \\
  350/  550/  350 & $3.20\cdot 10^{-4}$ & $6.43\cdot 10^{-4}$ & $9.35\cdot 10^{-2}$ & $4.51\cdot 10^{-1}$ & $4.11\cdot 10^{-1}$ & $9.55\cdot 10^{-1}$ \\
  350/  550/  375 & $3.20\cdot 10^{-4}$ & $6.42\cdot 10^{-4}$ & $9.33\cdot 10^{-2}$ & $4.51\cdot 10^{-1}$ & $4.10\cdot 10^{-1}$ & $9.55\cdot 10^{-1}$ \\
  350/  550/  400 & $3.19\cdot 10^{-4}$ & $6.41\cdot 10^{-4}$ & $9.32\cdot 10^{-2}$ & $4.51\cdot 10^{-1}$ & $4.10\cdot 10^{-1}$ & $9.54\cdot 10^{-1}$ \\
  600/  450/  300 & $2.52\cdot 10^{-4}$ & $5.05\cdot 10^{-4}$ & $6.92\cdot 10^{-2}$ & $3.29\cdot 10^{-1}$ & $3.02\cdot 10^{-1}$ & $7.00\cdot 10^{-1}$ \\
  600/  450/  350 & $2.53\cdot 10^{-4}$ & $5.07\cdot 10^{-4}$ & $6.96\cdot 10^{-2}$ & $3.31\cdot 10^{-1}$ & $3.04\cdot 10^{-1}$ & $7.04\cdot 10^{-1}$ \\
  600/  450/  400 & $2.52\cdot 10^{-4}$ & $5.05\cdot 10^{-4}$ & $6.92\cdot 10^{-2}$ & $3.30\cdot 10^{-1}$ & $3.02\cdot 10^{-1}$ & $7.01\cdot 10^{-1}$ \\
  600/  500/  300 & $2.47\cdot 10^{-4}$ & $4.95\cdot 10^{-4}$ & $6.77\cdot 10^{-2}$ & $3.24\cdot 10^{-1}$ & $2.96\cdot 10^{-1}$ & $6.88\cdot 10^{-1}$ \\
  600/  500/  350 & $2.48\cdot 10^{-4}$ & $4.97\cdot 10^{-4}$ & $6.81\cdot 10^{-2}$ & $3.26\cdot 10^{-1}$ & $2.98\cdot 10^{-1}$ & $6.92\cdot 10^{-1}$ \\
  600/  500/  400 & $2.47\cdot 10^{-4}$ & $4.94\cdot 10^{-4}$ & $6.77\cdot 10^{-2}$ & $3.24\cdot 10^{-1}$ & $2.96\cdot 10^{-1}$ & $6.88\cdot 10^{-1}$ \\
  600/  550/  300 & $2.40\cdot 10^{-4}$ & $4.81\cdot 10^{-4}$ & $6.58\cdot 10^{-2}$ & $3.16\cdot 10^{-1}$ & $2.89\cdot 10^{-1}$ & $6.71\cdot 10^{-1}$ \\
  600/  550/  350 & $2.41\cdot 10^{-4}$ & $4.83\cdot 10^{-4}$ & $6.63\cdot 10^{-2}$ & $3.18\cdot 10^{-1}$ & $2.91\cdot 10^{-1}$ & $6.75\cdot 10^{-1}$ \\
  600/  550/  400 & $2.40\cdot 10^{-4}$ & $4.80\cdot 10^{-4}$ & $6.58\cdot 10^{-2}$ & $3.17\cdot 10^{-1}$ & $2.89\cdot 10^{-1}$ & $6.71\cdot 10^{-1}$ \\
\hline\end{tabular}

%% file: plots/br_ff.tex
\begin{tabular}{|ccccccc|}
\hline $\vphantom{\bar{\bar{\bar{\PQb}}}}$%
\rule[-1ex]{0ex}{3.5ex}%
$\MH$ [GeV]&   $\Hbb$ & $\Htautau$ & $\Hmumu$ &   $\Hss$ &   $\Hcc$ &   $\Htt$ \\ \hline
\rule{0ex}{2.5ex}%
  100 & $5.70\cdot 10^{-1}$ & $5.98\cdot 10^{-2}$ & $2.08\cdot 10^{-4}$ & $2.44\cdot 10^{-4}$ & $2.88\cdot 10^{-2}$ & $0.00$ \\
  110 & $5.30\cdot 10^{-1}$ & $5.67\cdot 10^{-2}$ & $1.97\cdot 10^{-4}$ & $2.26\cdot 10^{-4}$ & $2.68\cdot 10^{-2}$ & $0.00$ \\
  120 & $4.87\cdot 10^{-1}$ & $5.29\cdot 10^{-2}$ & $1.84\cdot 10^{-4}$ & $2.08\cdot 10^{-4}$ & $2.46\cdot 10^{-2}$ & $0.00$ \\
  130 & $4.36\cdot 10^{-1}$ & $4.82\cdot 10^{-2}$ & $1.67\cdot 10^{-4}$ & $1.86\cdot 10^{-4}$ & $2.20\cdot 10^{-2}$ & $0.00$ \\
  140 & $3.72\cdot 10^{-1}$ & $4.17\cdot 10^{-2}$ & $1.45\cdot 10^{-4}$ & $1.59\cdot 10^{-4}$ & $1.88\cdot 10^{-2}$ & $0.00$ \\
  150 & $2.83\cdot 10^{-1}$ & $3.20\cdot 10^{-2}$ & $1.11\cdot 10^{-4}$ & $1.21\cdot 10^{-4}$ & $1.42\cdot 10^{-2}$ & $0.00$ \\
  160 & $1.13\cdot 10^{-1}$ & $1.29\cdot 10^{-2}$ & $4.48\cdot 10^{-5}$ & $4.80\cdot 10^{-5}$ & $5.67\cdot 10^{-3}$ & $0.00$ \\
  170 & $3.26\cdot 10^{-2}$ & $3.78\cdot 10^{-3}$ & $1.31\cdot 10^{-5}$ & $1.39\cdot 10^{-5}$ & $1.64\cdot 10^{-3}$ & $0.00$ \\
  180 & $2.15\cdot 10^{-2}$ & $2.52\cdot 10^{-3}$ & $8.74\cdot 10^{-6}$ & $9.17\cdot 10^{-6}$ & $1.08\cdot 10^{-3}$ & $0.00$ \\
  190 & $1.39\cdot 10^{-2}$ & $1.64\cdot 10^{-3}$ & $5.69\cdot 10^{-6}$ & $5.91\cdot 10^{-6}$ & $6.98\cdot 10^{-4}$ & $0.00$ \\
  200 & $1.06\cdot 10^{-2}$ & $1.27\cdot 10^{-3}$ & $4.41\cdot 10^{-6}$ & $4.53\cdot 10^{-6}$ & $5.35\cdot 10^{-4}$ & $0.00$ \\
  250 & $4.73\cdot 10^{-3}$ & $5.89\cdot 10^{-4}$ & $2.04\cdot 10^{-6}$ & $2.01\cdot 10^{-6}$ & $2.38\cdot 10^{-4}$ & $0.00$ \\
  300 & $2.70\cdot 10^{-3}$ & $3.48\cdot 10^{-4}$ & $1.21\cdot 10^{-6}$ & $1.15\cdot 10^{-6}$ & $1.36\cdot 10^{-4}$ & $3.26\cdot 10^{-4}$ \\
  400 & $6.38\cdot 10^{-4}$ & $8.63\cdot 10^{-5}$ & $2.99\cdot 10^{-7}$ & $2.71\cdot 10^{-7}$ & $3.20\cdot 10^{-5}$ & $4.50\cdot 10^{-1}$ \\
  500 & $2.96\cdot 10^{-4}$ & $4.16\cdot 10^{-5}$ & $1.44\cdot 10^{-7}$ & $1.26\cdot 10^{-7}$ & $1.48\cdot 10^{-5}$ & $5.22\cdot 10^{-1}$ \\
  600 & $2.02\cdot 10^{-4}$ & $2.92\cdot 10^{-5}$ & $1.01\cdot 10^{-7}$ & $8.58\cdot 10^{-8}$ & $1.01\cdot 10^{-5}$ & $4.82\cdot 10^{-1}$ \\
  700 & $1.52\cdot 10^{-4}$ & $2.26\cdot 10^{-5}$ & $7.82\cdot 10^{-8}$ & $6.46\cdot 10^{-8}$ & $7.61\cdot 10^{-6}$ & $4.21\cdot 10^{-1}$ \\
  800 & $1.18\cdot 10^{-4}$ & $1.80\cdot 10^{-5}$ & $6.24\cdot 10^{-8}$ & $5.02\cdot 10^{-8}$ & $5.91\cdot 10^{-6}$ & $3.56\cdot 10^{-1}$ \\
 1000 & $7.37\cdot 10^{-5}$ & $1.17\cdot 10^{-5}$ & $4.06\cdot 10^{-8}$ & $3.13\cdot 10^{-8}$ & $3.69\cdot 10^{-6}$ & $2.43\cdot 10^{-1}$%
\rule[-1ex]{0ex}{2.5ex}\\
\hline\end{tabular}


%% file: plots/br_vv.tex
\begin{tabular}{|cccccc|}
\hline
\rule[-1ex]{0ex}{3.5ex}
$\MH$ [GeV]&   $\Hgg$ & $\HZga$ &   $\HWW$ &   $\HZZ$ & $\Gamma_{\PH}$ [GeV] \\ \hline
  100 & $3.39\cdot 10^{-1}$ & $1.70\cdot 10^{-5}$ & $1.67\cdot 10^{-3}$ & $1.35\cdot 10^{-4}$ & $5.52\cdot 10^{-3}$ \\
  110 & $3.78\cdot 10^{-1}$ & $1.35\cdot 10^{-4}$ & $7.20\cdot 10^{-3}$ & $5.83\cdot 10^{-4}$ & $6.41\cdot 10^{-3}$ \\
  120 & $4.10\cdot 10^{-1}$ & $4.06\cdot 10^{-4}$ & $2.27\cdot 10^{-2}$ & $2.37\cdot 10^{-3}$ & $7.49\cdot 10^{-3}$ \\
  130 & $4.28\cdot 10^{-1}$ & $8.51\cdot 10^{-4}$ & $5.77\cdot 10^{-2}$ & $7.12\cdot 10^{-3}$ & $8.92\cdot 10^{-3}$ \\
  140 & $4.20\cdot 10^{-1}$ & $1.46\cdot 10^{-3}$ & $1.29\cdot 10^{-1}$ & $1.68\cdot 10^{-2}$ & $1.11\cdot 10^{-2}$ \\
  150 & $3.63\cdot 10^{-1}$ & $2.13\cdot 10^{-3}$ & $2.75\cdot 10^{-1}$ & $3.09\cdot 10^{-2}$ & $1.55\cdot 10^{-2}$ \\
  160 & $1.62\cdot 10^{-1}$ & $2.01\cdot 10^{-3}$ & $6.80\cdot 10^{-1}$ & $2.78\cdot 10^{-2}$ & $4.10\cdot 10^{-2}$ \\
  170 & $5.27\cdot 10^{-2}$ & $8.96\cdot 10^{-4}$ & $8.90\cdot 10^{-1}$ & $2.02\cdot 10^{-2}$ & $1.49\cdot 10^{-1}$ \\
  180 & $3.85\cdot 10^{-2}$ & $6.95\cdot 10^{-4}$ & $8.82\cdot 10^{-1}$ & $5.43\cdot 10^{-2}$ & $2.37\cdot 10^{-1}$ \\
  190 & $2.75\cdot 10^{-2}$ & $5.07\cdot 10^{-4}$ & $7.60\cdot 10^{-1}$ & $1.97\cdot 10^{-1}$ & $3.84\cdot 10^{-1}$ \\
  200 & $2.33\cdot 10^{-2}$ & $4.28\cdot 10^{-4}$ & $7.18\cdot 10^{-1}$ & $2.43\cdot 10^{-1}$ & $5.21\cdot 10^{-1}$ \\
  250 & $1.62\cdot 10^{-2}$ & $2.48\cdot 10^{-4}$ & $6.93\cdot 10^{-1}$ & $2.86\cdot 10^{-1}$ & $1.41$ \\
  300 & $1.39\cdot 10^{-2}$ & $1.58\cdot 10^{-4}$ & $6.84\cdot 10^{-1}$ & $3.00\cdot 10^{-1}$ & $2.87$ \\
  400 & $7.40\cdot 10^{-3}$ & $3.63\cdot 10^{-5}$ & $3.74\cdot 10^{-1}$ & $1.68\cdot 10^{-1}$ & $1.55\cdot 10^{1}$ \\
  500 & $4.28\cdot 10^{-3}$ & $1.41\cdot 10^{-5}$ & $3.22\cdot 10^{-1}$ & $1.51\cdot 10^{-1}$ & $4.06\cdot 10^{1}$ \\
  600 & $2.99\cdot 10^{-3}$ & $8.21\cdot 10^{-6}$ & $3.46\cdot 10^{-1}$ & $1.69\cdot 10^{-1}$ & $6.99\cdot 10^{1}$ \\
  700 & $2.17\cdot 10^{-3}$ & $5.46\cdot 10^{-6}$ & $3.86\cdot 10^{-1}$ & $1.91\cdot 10^{-1}$ & $1.06\cdot 10^{2}$ \\
  800 & $1.60\cdot 10^{-3}$ & $3.88\cdot 10^{-6}$ & $4.27\cdot 10^{-1}$ & $2.15\cdot 10^{-1}$ & $1.52\cdot 10^{2}$ \\
 1000 & $8.08\cdot 10^{-4}$ & $2.28\cdot 10^{-6}$ & $5.02\cdot 10^{-1}$ & $2.54\cdot 10^{-1}$ & $2.88\cdot 10^{2}$%
\rule[-1ex]{0ex}{2.5ex}\\
\hline\end{tabular}

%% file: plots/br_h4f_comb.tex
\begin{tabular}{|ccccccc|}
\hline
\rule[-1ex]{0ex}{3.5ex}%
$\MH$ [GeV]&   $\PH \to 4\Pe$ & $\PH \to 2\Pe 2\PGm$ &   $\PH \to 4\Pl$ &   $\PH \to 4\Pq$ & $\PH \to 2\Pl 2\Pq$ &   $\PH \to 4\Pf$ \\ \hline 
\rule[-0ex]{0ex}{2.5ex}%
   100 & $2.26\cdot 10^{-7}$ & $3.31\cdot 10^{-7}$ & $1.67\cdot 10^{-4}$ & $7.48\cdot 10^{-4}$ & $7.85\cdot 10^{-4}$ & $1.70\cdot 10^{-3}$ \\
   110 & $7.81\cdot 10^{-7}$ & $1.27\cdot 10^{-6}$ & $7.32\cdot 10^{-4}$ & $3.53\cdot 10^{-3}$ & $3.36\cdot 10^{-3}$ & $7.62\cdot 10^{-3}$ \\
   120 & $2.76\cdot 10^{-6}$ & $4.99\cdot 10^{-6}$ & $2.36\cdot 10^{-3}$ & $1.17\cdot 10^{-2}$ & $1.08\cdot 10^{-2}$ & $2.48\cdot 10^{-2}$ \\
   130 & $8.00\cdot 10^{-6}$ & $1.48\cdot 10^{-5}$ & $6.12\cdot 10^{-3}$ & $3.05\cdot 10^{-2}$ & $2.78\cdot 10^{-2}$ & $6.44\cdot 10^{-2}$ \\
   140 & $1.83\cdot 10^{-5}$ & $3.51\cdot 10^{-5}$ & $1.38\cdot 10^{-2}$ & $6.88\cdot 10^{-2}$ & $6.24\cdot 10^{-2}$ & $1.45\cdot 10^{-1}$ \\
   150 & $3.33\cdot 10^{-5}$ & $6.42\cdot 10^{-5}$ & $2.91\cdot 10^{-2}$ & $1.45\cdot 10^{-1}$ & $1.31\cdot 10^{-1}$ & $3.05\cdot 10^{-1}$ \\
   160 & $2.95\cdot 10^{-5}$ & $5.75\cdot 10^{-5}$ & $6.85\cdot 10^{-2}$ & $3.31\cdot 10^{-1}$ & $3.05\cdot 10^{-1}$ & $7.04\cdot 10^{-1}$ \\
   170 & $2.13\cdot 10^{-5}$ & $4.17\cdot 10^{-5}$ & $8.81\cdot 10^{-2}$ & $4.29\cdot 10^{-1}$ & $3.92\cdot 10^{-1}$ & $9.08\cdot 10^{-1}$ \\
   180 & $5.64\cdot 10^{-5}$ & $1.11\cdot 10^{-4}$ & $8.98\cdot 10^{-2}$ & $4.44\cdot 10^{-1}$ & $4.02\cdot 10^{-1}$ & $9.36\cdot 10^{-1}$ \\
   190 & $2.03\cdot 10^{-4}$ & $4.04\cdot 10^{-4}$ & $9.04\cdot 10^{-2}$ & $4.57\cdot 10^{-1}$ & $4.09\cdot 10^{-1}$ & $9.56\cdot 10^{-1}$ \\
   200 & $2.49\cdot 10^{-4}$ & $4.99\cdot 10^{-4}$ & $9.06\cdot 10^{-2}$ & $4.63\cdot 10^{-1}$ & $4.10\cdot 10^{-1}$ & $9.64\cdot 10^{-1}$ \\
   250 & $2.90\cdot 10^{-4}$ & $5.86\cdot 10^{-4}$ & $9.05\cdot 10^{-2}$ & $4.71\cdot 10^{-1}$ & $4.16\cdot 10^{-1}$ & $9.78\cdot 10^{-1}$ \\
   300 & $3.00\cdot 10^{-4}$ & $6.10\cdot 10^{-4}$ & $9.05\cdot 10^{-2}$ & $4.74\cdot 10^{-1}$ & $4.18\cdot 10^{-1}$ & $9.82\cdot 10^{-1}$ \\
   400 & $1.71\cdot 10^{-4}$ & $3.42\cdot 10^{-4}$ & $5.02\cdot 10^{-2}$ & $2.61\cdot 10^{-1}$ & $2.30\cdot 10^{-1}$ & $5.41\cdot 10^{-1}$ \\
   500 & $1.56\cdot 10^{-4}$ & $3.13\cdot 10^{-4}$ & $4.41\cdot 10^{-2}$ & $2.27\cdot 10^{-1}$ & $2.01\cdot 10^{-1}$ & $4.73\cdot 10^{-1}$ \\
   600 & $1.74\cdot 10^{-4}$ & $3.51\cdot 10^{-4}$ & $4.81\cdot 10^{-2}$ & $2.48\cdot 10^{-1}$ & $2.19\cdot 10^{-1}$ & $5.15\cdot 10^{-1}$ \\
   700 & $1.99\cdot 10^{-4}$ & $4.00\cdot 10^{-4}$ & $5.41\cdot 10^{-2}$ & $2.77\cdot 10^{-1}$ & $2.46\cdot 10^{-1}$ & $5.77\cdot 10^{-1}$ \\
   800 & $2.24\cdot 10^{-4}$ & $4.52\cdot 10^{-4}$ & $6.05\cdot 10^{-2}$ & $3.08\cdot 10^{-1}$ & $2.74\cdot 10^{-1}$ & $6.42\cdot 10^{-1}$ \\
  1000 & $2.70\cdot 10^{-4}$ & $5.42\cdot 10^{-4}$ & $7.21\cdot 10^{-2}$ & $3.60\cdot 10^{-1}$ & $3.23\cdot 10^{-1}$ & $7.56\cdot 10^{-1}$%
\rule[-1ex]{0ex}{2.5ex}\\
\hline\end{tabular}

%% file: plots/BR_gaga_scan_grid.tex
\begin{tabular}{|ccc|}
\hline
\rule[-1ex]{0ex}{3.5ex}%
$\MH\ /\ \Mbp\ /\ \Mnp$ &
w/o NLO EW       &
w/ NLO EW           \\{}
[GeV] & & \\
\hline
\rule[0ex]{0ex}{2.5ex}%
$120\ /\ 450\ /\ 350$ & $2.52\,\cdot\,10^{-4}$  & $9.91\,\cdot\,10^{-6}$ \\
$120\ /\ 450\ /\ 350$ & $2.52\,\cdot\,10^{-4}$  & $9.31\,\cdot\,10^{-6}$ \\
$120\ /\ 450\ /\ 350$ & $2.51\,\cdot\,10^{-4}$  & $8.69\,\cdot\,10^{-6}$ \\
\hline
$120\ /\ 500\ /\ 375$ & $2.50\,\cdot\,10^{-4}$  & $4.60\,\cdot\,10^{-6}$ \\
$120\ /\ 500\ /\ 375$ & $2.50\,\cdot\,10^{-4}$  & $4.20\,\cdot\,10^{-6}$ \\
$120\ /\ 500\ /\ 375$ & $2.49\,\cdot\,10^{-4}$  & $3.80\,\cdot\,10^{-6}$ \\
\hline
$120\ /\ 550\ /\ 400$ & $2.47\,\cdot\,10^{-4}$  & $1.13\,\cdot\,10^{-6}$ \\
$120\ /\ 550\ /\ 400$ & $2.47\,\cdot\,10^{-4}$  & $9.38\,\cdot\,10^{-7}$ \\
$120\ /\ 550\ /\ 400$ & $2.46\,\cdot\,10^{-4}$  & $7.56\,\cdot\,10^{-7}$%
\rule[-1ex]{0ex}{2.5ex}\\
\hline
\end{tabular}